\documentclass[
superscriptaddress,
aps,prl,
reprint,
twocolumn,
amsmath,amssymb,showpacs
]{revtex4-2}
\usepackage{newtxtext,newtxmath}
\usepackage{physics}
\usepackage{graphicx}
\usepackage{bbm}
\usepackage{amssymb,bm}
\usepackage{dsfont}
\usepackage{xcolor}
\usepackage{ulem} 
\definecolor{urlc}{RGB}{58,105,157}
\usepackage[
colorlinks=true,
urlcolor=urlc,
linkcolor=urlc,
citecolor=urlc
]{hyperref}

\begin{document}

\title{Correlation-enhanced metrology from scrambling dynamics in a solid-state spin system}

\author{Yu-Chen Li}
	% \thanks{These authors contribute equally to this work}
	\affiliation{Laboratory of Spin Magnetic Resonance, School of Physical Sciences,
		Anhui Province Key Laboratory of Scientific Instrument Development and Application,
		University of Science and Technology of China, Hefei 230026, China}
	
\author{Shengyu Zhang}
	% \thanks{These authors contribute equally to this work}
	\affiliation{Laboratory of Spin Magnetic Resonance, School of Physical Sciences,
		Anhui Province Key Laboratory of Scientific Instrument Development and Application,
		University of Science and Technology of China, Hefei 230026, China}
	\affiliation{Hefei National Laboratory, Hefei 230088, China}
	
\author{Ze Wu}
	\affiliation{Department of Physics, The Chinese University of Hong Kong, Hong Kong, China}
	\affiliation{Laboratory of Spin Magnetic Resonance, School of Physical Sciences,
		Anhui Province Key Laboratory of Scientific Instrument Development and Application,
		University of Science and Technology of China, Hefei 230026, China}

	\author{Haochuan Yin}
	\affiliation{Laboratory of Spin Magnetic Resonance, School of Physical Sciences,
		Anhui Province Key Laboratory of Scientific Instrument Development and Application,
		University of Science and Technology of China, Hefei 230026, China}
	\affiliation{Hefei National Laboratory, Hefei 230088, China}
    
	\author{Liqiang Zhao}
	\affiliation{Laboratory of Spin Magnetic Resonance, School of Physical Sciences,
		Anhui Province Key Laboratory of Scientific Instrument Development and Application,
		University of Science and Technology of China, Hefei 230026, China}
	\affiliation{Hefei National Laboratory, Hefei 230088, China}

    \author{Xiaoxue An}
	\affiliation{Laboratory of Spin Magnetic Resonance, School of Physical Sciences,
		Anhui Province Key Laboratory of Scientific Instrument Development and Application,
		University of Science and Technology of China, Hefei 230026, China}
	\affiliation{Hefei National Research Center for Physical Sciences at the Microscale, Hefei 230026, China}

    \author{Jiaxi Cui}
	\affiliation{Laboratory of Spin Magnetic Resonance, School of Physical Sciences,
		Anhui Province Key Laboratory of Scientific Instrument Development and Application,
		University of Science and Technology of China, Hefei 230026, China}
	\affiliation{Hefei National Research Center for Physical Sciences at the Microscale, Hefei 230026, China}
    
	\author{Dieter Suter}
    \email{Dieter.Suter@tu-dortmund.de}
    \affiliation{Laboratory of Spin Magnetic Resonance, School of Physical Sciences,
		Anhui Province Key Laboratory of Scientific Instrument Development and Application,
		University of Science and Technology of China, Hefei 230026, China}
    \affiliation{Fakultät Physik, Technische Universität Dortmund, Dortmund D-44221, Germany}
    
	\author{Xinhua Peng}
	\email{xhpeng@ustc.edu.cn}
	\affiliation{Laboratory of Spin Magnetic Resonance, School of Physical Sciences,
		Anhui Province Key Laboratory of Scientific Instrument Development and Application,
		University of Science and Technology of China, Hefei 230026, China}
 	\affiliation{Hefei National Laboratory, Hefei 230088, China}
    \affiliation{Hefei National Research Center for Physical Sciences at the Microscale, Hefei 230026, China}

\date{\today}

\begin{abstract}
Quantum information scrambling, the dispersal of local information into many-body degrees of freedom, provides a powerful mechanism for generating large-scale correlations and entanglement essential for quantum-enhanced metrology. However, experimentally verifying such quantum-enhanced metrology remains a demanding task. Here, we correlate thousands of spins by engineering chaotic scrambling dynamics in a solid-state nuclear spin system. By leveraging the newly developed scramblon theory, we reveal exponential scaling in both the quantum Fisher information and the signal response to a phase shift. The signal response achieves a correlation-enabled enhancement of $33(2)$ dB over uncorrelated spins. After accounting for signal loss due to imperfect time reversal in the readout stage, we obtain a total metrological gain of 18(1) dB with a phase sensitivity of 40(3) ${\mathrm{\mu rad}}$. Our results bridge quantum chaos with practical quantum metrology, establishing reversible scrambling dynamics as a powerful resource for precision measurements.
\end{abstract}

\maketitle

Quantum information scrambling is critical for understanding quantum chaos and thermalization in isolated quantum systems~\cite{DeutschQuantumStatisticalMechanics1991,SrednickiApproachThermalEquilibrium1999,Rigol2007ThermalizationAI,Kaufman2016QuantumTT,Swingle:2018ekw}. Over the past decade, the out-of-time-ordered correlator (OTOC) has emerged as the primary diagnostic of scrambling and many-body chaos~\cite{larkin1969quasiclassical,kitaev2014hidden,shenkerBlackHolesButterfly2014,robertsLocalizedShocks2015,shenkerStringyEffectsScrambling2015,Maldacena_2016}, with its early experimental observations reported in nuclear magnetic resonance (NMR) \cite{DuMeasuringOutofTimeOrderCorrelators2017a} and trapped-ion \cite{ReyMeasuringOutoftimeorderCorrelations2017a} systems. Since then, OTOC measurements have been extensively validated across diverse platforms, including cold atoms~\cite{VuleticTimereversalbasedQuantumMetrology2022,VuleticImprovingMetrologyQuantum2023,ChinQuantumSimulationUnruh2019,ThomasEnergyResolvedInformationScrambling2021,LiObservationQuantumInformation2024a,YouObservationAnomalousInformation2024a,WeidemullerTimereversalDipolarQuantum2024}, superconducting qubits~\cite{ChenInformationScramblingQuantum2021,DuanInformationScramblingDynamics2022,OliverProbingQuantumInformation2022,ZhaoProbingOperatorSpreading2022,ZobristConstructiveInterferenceEdge2025,ge2025information,hu2026quantum}, nitrogen vacancy centers~\cite{gao2025signal}, and also continuously in trapped ions~\cite{MonroeVerifiedQuantumInformation2019,ReyUnifyingScramblingThermalization2019b,LinkeExperimentalMeasurementOutofTimeOrdered2022} and NMR systems~\cite{CappellaroExploringLocalizationNuclear2018a,CappellaroEmergentPrethermalizationSignatures2019,LiExperimentalObservationEquilibrium2020,sanchez2020perturbation,dominguez2021decoherence,PastawskiEmergentDecoherenceInduced2022,DuEmergentUniversalQuench2024a,li2026errorresilient}. Despite these efforts, exponential scrambling—characterized by a Lyapunov exponent—has remained notoriously difficult to observe directly in many-body experiments, owing to the stringent demands on time-reversal fidelity and signal stability. Recently, the scramblon theory was proposed to provide a universal analytical framework for chaotic scrambling dynamics \cite{kitaev2018soft,gu2019relation,stanford2022subleading,gu2022two,Liu2023sig}. Building on this framework, we have experimentally uncovered, for the first time, the anticipated exponential growth of quantum many-body chaos and extracted the quantum Lyapunov exponent in a solid-state NMR system with thousands of correlated nuclear spins~\cite{li2026errorresilient}. Beyond its fundamental interest, scrambling dynamics offers a powerful resource for quantum metrology, enabling entanglement-enhanced metrology over $N$ independent probes~\cite{VuleticImprovingMetrologyQuantum2023,kobrin2024universal,ge2025information,hu2026quantum}. Although such protocols have been experimentally verified on cold-atom~\cite{VuleticImprovingMetrologyQuantum2023} and superconducting-qubit~\cite{ge2025information,hu2026quantum} platforms, these prior experiments involved only a limited number of particles (up to 200), and an extension to many-body chaotic systems remains an outstanding challenge.

In this work, we apply the scramblon theory to the framework of quantum-enhanced sensing. Sensing tasks are generally implemented by applying an observable (e.g., angular momentum) for some time such that it changes the phases of some density operator elements, and the goal is to optimize the sensitivity, i.e., the resolvable phase change. By fitting the scramblon ansatz to experimental OTOC data, we successfully predict and verify an initial exponential amplification of the phase sensitivity, characterized by the Lyapunov exponent, from thousands of correlated spins under imperfect time-reversal readout. Leveraging the mathematical relation between the quantum Fisher information (QFI) and the OTOC, we demonstrate that the QFI also increases exponentially with the same Lyapunov exponent, which quantifies the ultimate precision limit for parameter estimation and, equivalently, the metrologically useful correlations. Our results establish a direct, theory-guided route from the verified observation of quantum chaos to practical quantum-enhanced metrology, highlighting the potential of reversible scrambling dynamics as a resource for high-precision measurements.

\begin{figure}
\centering
\includegraphics[scale=1.1]{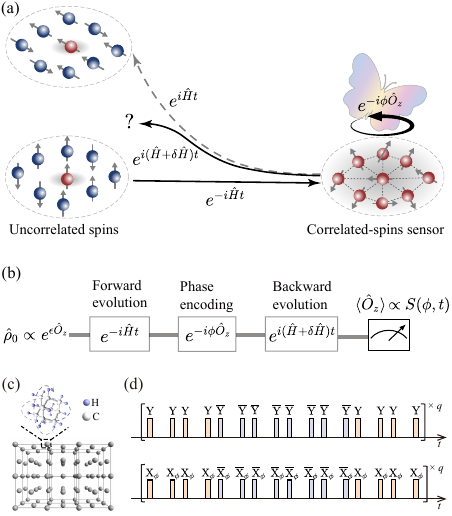}
\caption{Schematic of the time-reversal protocol. (a) Initially uncorrelated spins develop multi-body correlations during scrambling dynamics governed by the Hamiltonian $\hat{H}$. The system is then subjected to a global rotation with phase $\phi$, followed by a time-reversed evolution that maps the encoded non-local information back onto local observables, yielding an amplified signal dependence on $\phi$. Under ideal, perfectly reversed evolution, the phase sensitivity exhibits an exponential growth, resembling the butterfly effect. However, in the presence of inevitable time-reversal imperfections $\delta\hat{H}$, this sensitivity is severely compromised. (b) Experimental sequence for MQC and OTOC measurements. (c) Adamantane molecules are arranged in a face-centered cubic lattice, where each molecule consists of spin-$1/2$ $^1$H nuclei and spinless $^{12}$C atoms. (d) Top: The Floquet pulse sequence that converts the secular dipolar Hamiltonian into a form dominated by the double-quantum-transition Hamiltonian $\hat{H}_{\rm dq}$, as defined by Eq.~\eqref{eq:dq_Ham}~\cite{yen1983multiple,baum1985multiple}. Bottom: The corresponding pulse sequence that implements the global rotation $e^{-i\phi\hat{O}_z}$, followed by backward evolution dominated by $-\hat{H}_{\rm dq}$ (see Supplementary Material \cite{SM} for details).
}
\label{fig1}
\end{figure}

\textit{Experimental protocol}\textemdash We experimentally implement the time-reversal protocol—a well-established method for measuring the multiple quantum coherence (MQC) spectrum and the OTOC~\cite{yen1983multiple,baum1985multiple,ReyMeasuringOutoftimeorderCorrelations2017a}—using hydrogen nuclear spins in a powdered adamantane (${\rm C_{10}H_{16}}$) sample (Fig.~\ref{fig1}). Each adamantane molecule contains 16 spin-$1/2$ hydrogen nuclei, denoted by $\bm{S}_{i\mu}$, where the index $i$ labels the molecule and $\mu=1,\dots,16$ identifies the individual spin. Due to rapid molecular thermal tumbling, intramolecular spin interactions are effectively averaged out, leaving the intermolecular dipolar coupling as the dominant spin-spin interaction. The sample is placed in a $9.4\text{-T}$ magnetic field at room temperature, ensuring that internal nuclear interactions remain perturbative compared to the dominant Zeeman energy. In the rotating frame at the Zeeman frequency, the system is governed by the secular dipolar Hamiltonian $\hat{H}_{\mathrm{dd}} \equiv \sum_{i<j,\mu,\nu} J_{ij} ( -\hat{S}^x_{i\mu}\hat{S}^x_{j\nu} - \hat{S}^y_{i\mu}\hat{S}^y_{j\nu} + 2\hat{S}^z_{i\mu}\hat{S}^z_{j\nu})$~\cite{abragam1961}. The intermolecular coupling $J_{ij}$ scales as $(1-3\cos^2\theta_{ij})/R_{ij}^3$, where $R_{ij}$ denotes the distance between molecular centers and $\theta_{ij}$ is the angle between $\bm{R}_{ij}$ and the external magnetic field along the $z$-axis. In an ensemble of randomly oriented crystal grains, the couplings $J_{ij}$ exhibit the behavior of random variables with a zero mean ($\overline{J_{ij}}=0$) and a variance of $\overline{J_{ij}^2}=4J^2/N$, where $N$ is the total number of spins and $J/(2\pi)\approx 1460$ Hz is determined via experimental calibration \cite{DuEmergentUniversalQuench2024a,li2026errorresilient}.

We prepare the system in an uncorrelated initial state weakly polarized along the $\hat{z}$ direction, $\hat{\rho}_{0}={\rm e}^{\epsilon\hat{O}_z}/\mathcal{Z}$, where $\epsilon\sim 10^{-5}$, $\mathcal{Z}$ is the partition function and $\hat{O}_z=\sum_{i\mu}^N\hat{S}_{i\mu}^{z}$ denotes the collective spin operator. We can omit the higher-order terms ${\rm O}(\epsilon^2)$ and consider an initial density matrix $\hat{\rho}_0=(\mathbbm{1}+\epsilon\hat{O}_z)/2^N$. Since the secular dipolar Hamiltonian commutes with $\hat{O}_z$, we engineer a Hamiltonian $\hat{H}$ satisfying $[\hat{H},\hat{\rho}_0]\ne 0$ to drive the system evolution and establish spin-spin correlations. Via periodic radio-frequency pulses in the regime of average Hamiltonian theory, we engineer $\hat{H}_{\mathrm{dd}}$ into an effective Hamiltonian $\hat{H}=\sum_n\hat{H}_{\rm F}^{(n)}$, described by the Floquet-Magnus expansion \cite{blanes2009magnus} (see Supplementary Material \cite{SM} for details). The strength of the $n$-th order term is on the order of $\mathrm{O}[(JT)^2]$ with $T$ the period of the pulse sequence. Except for limited cases constrained by conservation laws such as $U(1)$ symmetry, the engineered time evolution is generally chaotic in 3D solid-state spin systems and the spins undergo quantum scrambling to develop complex multi-body correlations that spread information across the ensemble~\cite{KaiserLocalizationdelocalizationTransitionDynamics2015,zhou2023operator,li2026errorresilient,jyoti2017dipolarly}. We use a widely adopted pulse sequence from Refs.~\cite{yen1983multiple,baum1985multiple} with $T=96\ \mu$s [Fig.~\ref{fig1}(d), top], which creates the leading-order component $\hat{H}^\mathrm{(0)}_{\rm F}$ as the double-quantum-transition Hamiltonian
\begin{equation}
\hat{H}\approx\hat{H}_{\mathrm{dq}} \equiv \sum\limits_{i<j,\mu,\nu}\frac{J_{ij}}{2}(\hat{S}_{i\mu}^+\hat{S}_{j\nu}^++\hat{S}_{i\mu}^-\hat{S}_{j\nu}^-).
\label{eq:dq_Ham}
\end{equation}

As illustrated in Fig.~\ref{fig1}(a), after an evolution time $t$, the correlated state $\hat{\rho}_t$ is created. Then a global phase shift $\phi$ is imprinted via the operation $e^{-i\phi\hat{O}_z}$, and the subsequent time-reversed evolution $e^{iHt}$ refocuses the delocalized information back onto local spin observables. To achieve this, a standard approach is to introduce a $\pi/2+\phi$ phase shift to the original pulse sequence, thereby engineering a phase-dependent Hamiltonian $e^{i\hat{O}_z(\frac{\pi}{2}+\phi)}\hat{H}e^{-i\hat{O}_z(\frac{\pi}{2}+\phi)}\approx e^{i\hat{O}_z\phi}(-\hat{H}_{\mathrm{dq}})e^{-i\hat{O}_z\phi}$ [Fig.~\ref{fig1}(d), bottom] (see Supplementary Material \cite{SM} for details). The combined forward and phase-shifted backward evolution achieves a total evolution $\hat{U}(\phi,t)=e^{i\hat{O}_z\phi}e^{i\hat{H}t}e^{-i\hat{O}_z\phi}e^{-i\hat{H}t}$. Furthermore, we choose the final observable to be $\hat{O}_z$ so that the last operator $e^{i\hat{O}_z\phi}$ has no effect and can be omitted. The final signal (after normalization) reduces to
\begin{equation}
\begin{aligned}
        S(\phi,t)
        &=\frac{1}{\mathcal{C}}\text{Tr}[ \hat{O}_z(t){e}^{- i\phi\hat{O}_z}\hat{O}_z(t){e}^{ i\phi\hat{O}_z}]\\
        &=\sum_m I_m(t) \cos(m\phi),
\end{aligned}
\end{equation}
which is an OTOC between $\hat{O}_z(t)=e^{-i\hat{H}t}\hat{O}_ze^{i\hat{H}t}$ and $e^{-i\phi\hat{O}_z}$, and can be decomposed into the MQC spectrum. Here, $I_m(t)$ denotes the MQC intensity of order $m$ within the operator $\hat{O}_z(t)$~\cite{yen1983multiple,baum1985multiple}. As the evolution time $t$ increases, the MQCs $I_m$ spread to higher orders, leading to amplified phase dependence. The normalization coefficient is $\mathcal{C}=\text{Tr}(\hat{O}_z^2)$, ensuring $\sum_m I_m(t)=S(0,t)=1$. Therefore, the phase sensitivity $\Delta\phi\equiv\Delta S|\partial_\phi S|^{-1}$, depends on the readout noise $\Delta S$ and the derivative $|\partial_\phi S|=\sum_m mI_m(t)\sin m\phi$. For a fixed evolution time $t$, there exists an optimal bias phase $\phi_{\rm opt}(t)$ at which $|\partial_\phi S(\phi,t)|$ is maximized, and we define $g(t)\equiv\max_{\phi}|\partial_\phi S(\phi,t)|$ as the signal response. An uncorrelated probe state $\hat{\rho}_{(\rm uncorre)} \propto \mathbbm{1} + \epsilon \hat{O}_x$, undergoing $e^{-i\phi\hat{O}_z}$ and then measured for $\langle \hat{O}_x\rangle$, gives a normalized signal $S_{(\rm uncorre)}=\cos\phi$ and $g_{(\rm uncorre)}=\max(|\sin\phi|)=1$. 

The distribution of $I_m$ can be well approximated by a Gaussian profile, $I_m\propto e^{-m^2/K}$, where the variance $K=2\sum_m m^2 I_m$ is referred to as the cluster size, quantifying the effective number of correlated spins~\cite{baum1985multiple,KaiserLocalizationdelocalizationTransitionDynamics2015}. This Gaussian approximation is physically grounded in the assumption that all quantum coherences (the off-diagonal elements of the density matrix) within the cluster are populated with equal probability during the quantum scrambling dynamics. As detailed in the End Matter, we derive that the QFI of the evolved state $\hat{\rho}_t = e^{-i\hat{H}t}\hat{\rho}_0 e^{i\hat{H}t}$ subject to a global rotation $e^{i\phi\hat{O}_z}$ can be accurately approximated by the second moment of the MQC spectrum:
\begin{equation}
\mathcal{F}_{\rm Q}(\hat{\rho}_t,\hat{O}_z)\approx\frac{N}{4}\epsilon^2 \sum_m m^2 I_m(t).
\label{eq:QFI}
\end{equation}
For an uncorrelated probe state $\hat{\rho}_{(\rm uncorre)} \propto \mathbbm{1} + \epsilon \hat{O}_x$, which contains only single-quantum coherences ($I_{\pm1} = 1/2$), the QFI is $\mathcal{F}_{\rm Q}(0) \approx \frac{N}{4}\epsilon^2$ \cite{modi2011quantum}. This uncorrelated probe state is prepared by applying a single $\pi/2$ pulse to $\hat{\rho}_0$. To quantify the metrological advantage gained from using correlated spin clusters, we define the QFI gain as $\mathcal{G}_{\mathrm{Q}}(t) \equiv \mathcal{F}_{\rm Q}(t)/\mathcal{F}_{\rm Q}(0) \approx \sum_m m^2 I_m(t)=K(t)/2=|\partial_\phi^2 S|_{\phi=0}$. In addition, the equality $|\partial_\phi^2 S|_{\phi=0}=-\frac{1}{\mathcal{C}}{\rm Tr}([\hat{O}_z(t),\hat{O}_z]^2)$ connects the QFI gain to the OTO commutator, which characterizes the growth of the operator size of $\hat{O}_z(t)$ under Heisenberg evolution during the scrambling dynamics~\cite{garttner2018relating,CappellaroEmergentPrethermalizationSignatures2019,zhou2023operator,li2026errorresilient}. 

\begin{figure}
\centering
\includegraphics[scale=0.9]{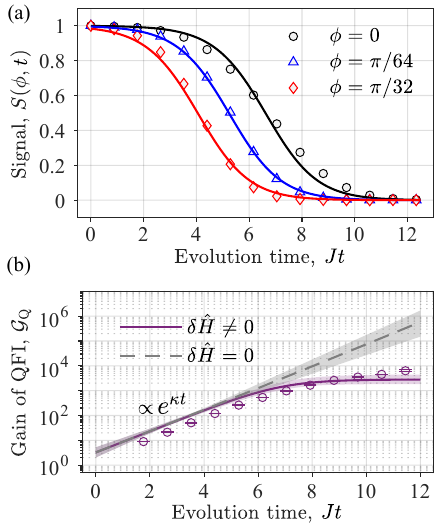}
\caption{Application of the scramblon ansatz and characterization of the QFI. (a) Fitting of the OTOC data using Eq.~\eqref{fitting}. The readout noise $\Delta S$ is on the order of $10^{-4}$ so that the error bars are contained within the data markers. The extracted parameters $a=0.0012(8)$, $b=1.5(6)$, $\Delta=0.6(3)$, and $\varkappa/J=1.0(2)$ are obtained via bootstrap fitting~\cite{hardle1991bootstrap}. (b) Setting $a=0$ in Eq.~\eqref{fitting} extrapolates the OTOC to its error-free limit, yielding the QFI gain and the effective number of correlated spins $K$, where $\mathcal{G}_{\rm Q}=K(t)/2=|\partial_\phi^2 S|_{\phi=0}=4b\Delta e^{\varkappa t}$ (gray dashed line). For comparison, estimates of $K(t)$ obtained by Gaussian fitting the FFT spectra of the re-normalized signal $\tilde{S}(\phi,t)=S(\phi, t)/S(0,t)$ are also plotted (purple spheres), alongside the theoretical curve $|\partial_\phi^2 \tilde{S}|_{\phi=0}=4b\Delta {\rm e}^{\varkappa t}/(1+ae^{\varkappa t})$ (purple solid curve). Shaded regions represent the 95\% confidence intervals derived from the parameter distributions generated via bootstrapping.
}
\label{fig2}
\end{figure}

\textit{Application of the scramblon ansatz for imperfect time reversal}\textemdash In the above discussion we have assumed that the time-reversal process is perfect, but this is never the case in real experiments [Fig.~\ref{fig1}(b)]. This is not only because pulse errors and decoherence are inevitable; more importantly, the higher-order average Hamiltonian terms $\sum_{n>0}\hat{H}_{\rm F}^{(n)}$ in the Floquet-Magnus expansion generally cannot be simultaneously reversed with the zeroth-order term $\hat{H}_{\rm F}^{(0)}$. We define $\delta\hat{H}$ as the sum of the two effective Hamiltonians governing the forward and backward evolutions. In the ideal case $\delta\hat{H}=0$, but when $\delta\hat{H}\ne 0$, the final signal becomes a more complicated correlation function:
\begin{equation}
		S(\phi,t)=\frac{1}{\mathcal{C}}\text{Tr}[ {\rm e}^{-{\rm i}\phi\hat{O}_z}\hat{O}_z(t){\rm e}^{{\rm i}\phi\hat{O}_z}\hat{\mathcal{V}}(t)\hat{O}_z(t)\hat{\mathcal{V}}^\dagger(t)], \label{Sphit}
\end{equation}
where $\hat{\mathcal{V}}(t) \equiv \mathbf{T}\exp\left({\rm i}\int_0^t{\rm d} t'\delta\hat{H}(t')\right)$ represents the time-evolution operator in the interaction picture, with $\mathbf{T}$ denoting time ordering. In this case, $S(0, t)$ is not constant but instead decays as $t$ increases, as shown by the gray open circles in Fig.~\ref{fig2}(a). This signal, also known as the Loschmidt echo, directly probes the fidelity of the reversed dynamics \cite{jalabert2001environment,gorin2006dynamics,macri2016loschmidt,yan2020information,sanchez2020perturbation,liu2025variational}. To address this, we employ the recently developed scramblon theory, which quantifies the complex interplay among the time-evolved operators in Eq.~\eqref{Sphit} \cite{gu2022two,Liu2023sig}. Importantly, the scramblon-theory-guided analysis of experimental data quantifies the competition between the metrological loss due to time-reversal imperfection and the sensitivity gain from intrinsic quantum chaos. The scramblons are postulated as collective excitations mediating the scrambling dynamics, but unlike the phonons or spin waves, which only damp or oscillate, scramblon modes can display growth characterized by the Lyapunov exponents. By assuming that only the mode with the largest Lyapunov exponent is dominant, this theory provides a universal ansatz for $S(\phi, t)$ in the limit of small $\phi$ \cite{li2026errorresilient}:
\begin{equation}
S(\phi,t)=\frac{1}{\left(1+a {\rm e}^{\varkappa t}+b \phi^2 {\rm e}^{\varkappa t}\right)^{2\Delta}},\label{fitting}
\end{equation}
where $a, b, \Delta,$ and $\varkappa$ are fitting parameters. Here, $a \neq 0$ quantifies the degree of time-reversal imperfection, while $\varkappa$ denotes the dominant Lyapunov exponent. We measure the OTOC data for $\phi=0, \pi/64,$ and $\pi/32$, and fit them using this ansatz [solid curves in Fig.~\ref{fig2}(a)]. The fitted results yield $a=0.0012(8)$, $b=1.5(6)$, $\Delta=0.6(3)$, and $\varkappa/J=1.0(2)$. 

\begin{figure*}
    \centering
    \includegraphics[scale=0.9]{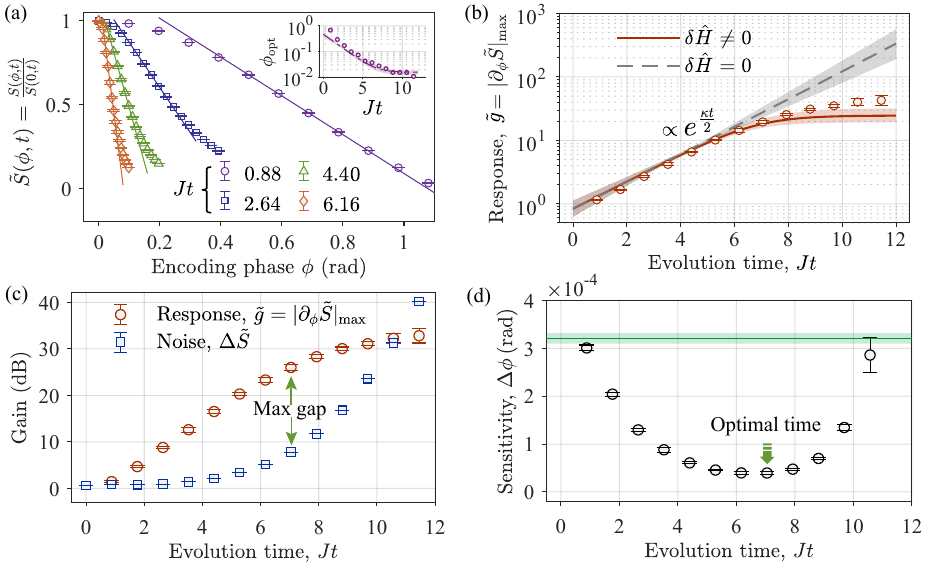}
    \caption{Signal response and phase sensitivity. (a) Re-normalized signal $\tilde{S}(\phi,t)=S(\phi,t)/S(0,t)$. The signal response $\tilde{g}=|\partial_\phi\tilde{S}|_{\max}$ and the optimal bias phase $\phi_{\rm opt}$ are extracted via linear fits using a 5-point sliding window. Data points represent the average of 10 repeated measurements, with error bars indicating the standard deviation. Inset: Time dependence of the optimal bias phase $\phi_{\rm opt}$, which reaches a plateau of $0.014(3)$, consistent with the theoretical expectation of $0.016(4)$ derived from Eq.~\eqref{eq:phi_opt} (shaded dashed line). (b) Extracted response $\tilde{g}$ as a function of evolution time, compared with theoretical predictions in the presence ($a \neq 0$, red solid curve) and absence ($a = 0$, gray dashed line) of time-reversal imperfections [Eq.~\eqref{eq:response}]. (c) Gains of the signal response $\tilde{g}=|\partial_\phi\tilde{S}|_{\max}$ and the noise-to-signal ratio $\Delta\tilde{S}$ relative to the uncorrelated probe state $\hat{\rho}_{(\rm uncorre)}\propto \mathbbm{1}+\epsilon\hat{O}_x$. (d) Phase sensitivity $[\Delta\phi]_{\min}=\Delta\tilde{S}|\partial_\phi\tilde{S}|_{\max}^{-1}$. The optimal sensitivity achieved by the correlated spins is 40(3) $\mu$rad at $Jt=7.05$, representing an 18(1) dB improvement over that of the uncorrelated spins (321(11) $\mu$rad, green shaded line). Error bars in (b)–(d) and all shaded regions denote the 95\% confidence intervals.
    }
    \label{fig3}
\end{figure*}

To account for the decay of the Loschmidt echo $S(0,t)$, it is standard practice in the literature to employ the re-normalized signal 
\begin{equation}
\tilde{S}(\phi,t) =\frac{S(\phi, t)}{S(0,t)}=\left(\frac{1+a {\rm e}^{\varkappa t}}{1+a {\rm e}^{\varkappa t}+b \phi^2 {\rm e}^{\varkappa t}}\right)^{2\Delta},
\label{eq:S_tilde}
\end{equation}
which ensures $\tilde{S}(0,t)=1$~\cite{ReyMeasuringOutoftimeorderCorrelations2017a,CappellaroEmergentPrethermalizationSignatures2019,ChenInformationScramblingQuantum2021,hu2026quantum}. From this, we obtain the analytical form for the QFI gain:
\begin{equation}
\mathcal{G}_{\mathrm{Q}}(t) =|\partial_\phi^2 \tilde{S}|_{\phi=0}= \frac{4b\Delta {\rm e}^{\varkappa t}}{1+ae^{\varkappa t}}.
\label{eq:gain_QFI}
\end{equation}
As illustrated in Fig.~\ref{fig2}(b), the QFI is predicted to grow exponentially as $4b\Delta {\rm e}^{\varkappa t}$ (gray dashed line) in the error-free limit $a=0$ ($\delta\hat{H}=0$). For the non-ideal case $\delta\hat{H}\ne0$, it deviates from exponential growth when $t\gtrsim -\frac{\log a}{\varkappa}$ and saturates to $4b\Delta/a$ (purple solid curve). On the experimental side, this signal can be expanded as $\tilde{S}(\phi,t) = \sum_m A_m(t) \mathrm{e}^{-\mathrm{i} m \phi}$, where the coefficients $A_m(t)$ correspond to the MQC intensities $I_m(t)$ in the ideal limit $\delta\hat{H} = 0$. By sampling $\phi$ from $0$ to $2\pi$ and performing a Fourier transform of $\tilde{S}(\phi,t)$ with respect to $\phi$, the spectrum of $A_m$ is extracted (see Supplementary Material \cite{SM} for details). The effective number of correlated spins, $K(t)$, is then determined via Gaussian fits to these spectra and plotted as purple circles in Fig.~\ref{fig2}(b), aligning closely with the theoretical prediction Eq.~\eqref{eq:gain_QFI}. It should be noted that this formula breaks down in the short-time limit ($Jt \ll 1$), where the single-scramblon approximation is no longer valid. Experimentally, the Gaussian approximation for spin clusters also fails in this early-time regime.

\textit{Calibration of phase sensitivity}\textemdash Having quantified the QFI, we now assess the phase sensitivity $\Delta\phi\equiv\Delta\tilde{S}|\partial_\phi\tilde{S}|^{-1}$. The phase dependence of $\tilde{S}(\phi,t)$ at various evolution times $Jt$ is presented in Fig.~\ref{fig3}(a). We perform linear fits to $\tilde{S}(\phi,t)$ using a 5-point sliding window, identifying the maximum slope as the signal response $\tilde{g}(t)\equiv\max_{\phi}|\partial_\phi\tilde{S}(\phi,t)|$, and the corresponding phase as the optimal bias phase $\phi_{\rm opt}(t)$. As the evolution time $t$ increases, $\phi_{\rm opt}(t)$ decreases and reaches a plateau at $\overline{\phi_{\rm opt}}=0.014(3)$, calculated as the mean and standard deviation for data in the range $Jt \in [8.81, 11.45]$. This result is consistent with the theoretical value obtained by maximizing $|\partial_\phi\tilde{S}|$ from Eq.~\eqref{eq:S_tilde} [see inset in Fig.~\ref{fig3}(a)]:
\begin{equation}
\lim_{t\to \infty}\phi_{\rm opt}(t)= \lim_{t\to \infty}\sqrt{\frac{1+ae^{\varkappa t}}{be^{\varkappa t}(4\Delta+1)}}=0.016(4).
\label{eq:phi_opt}
\end{equation}
Accordingly, the maximum slope follows the analytical expression:
\begin{equation}
\tilde{g}(t)\equiv|\partial_\phi\tilde{S}|_{\max}=D\sqrt{\frac{b e^{\varkappa t}}{1+a e^{\varkappa t}}},
\label{eq:response}
\end{equation}
where $D=\frac{2\Delta(4\Delta+1)^{2\Delta+1/2}}{2^{2\Delta}(2\Delta+1)^{2\Delta+1}}$. This formula predicts an initial exponential growth with exponent $\varkappa/2$, followed by a saturation behavior depending on the presence ($a \neq 0$) or absence ($a = 0$) of time-reversal imperfections [shaded curves in Fig.~\ref{fig3}(b)]. The former case ($a \neq 0$) is in good agreement with our experimental observations [red spheres in Fig.~\ref{fig3}(b)].

For comparison, the signal response of an uncorrelated probe state, $\hat{\rho}_{(\rm uncorre)}\propto\mathbbm{1}+\epsilon\hat{O}_x$, is measured as $\tilde{g}_{(\rm uncorre)}=   0.98(3)$, consistent with the theoretical value $\max(|\sin\phi|)=1$ (see Supplementary Material \cite{SM} for details). Consequently, the gain $\tilde{g}(t)/\tilde{g}_{(\rm uncorre)}$—reaching a maximum of $33(2)$ dB at $Jt=11.45$—is shown by the red spheres in Fig.~\ref{fig3}(c). The re-normalized readout noise is given by $\Delta\tilde{S}=\Delta S/S(0,t)$, where $\Delta S$ is the standard error contributed by the circuit noise of the nuclear spin FID detection, which is independent of $t$ and $\phi$ (see Supplementary Material \cite{SM} for details). However, as the Loschmidt echo $S(0,t)$ decays, $\Delta\tilde{S}$ increases from $3.4(1)\times 10^{-4}$ at $Jt=0$ to 0.032(1) at $Jt=11.45$. For the uncorrelated state, $\Delta\tilde{S}_{(\rm uncorre)}=3.1(1)\times 10^{-4}$. The increase of readout noise $\Delta\tilde{S}/\Delta\tilde{S}_{(\rm uncorre)}$ is plotted as blue squares in Fig.~\ref{fig3}(c). This competition between the amplification of signal response and the increase of noise yields an optimal evolution time at which the net metrological gain—the difference between the response gain and noise increase—is maximized. As shown in Fig.~\ref{fig3}(d), the phase sensitivity $\Delta\phi$ reaches an optimal value of 40(3) $\mu$rad at $Jt=7.05$ with a bias phase of $\phi_{\rm opt}=0.028$ rad, representing an 18(1) dB improvement over the sensitivity of uncorrelated spins [321(11) $\mu$rad].

\textit{Conclusion}\textemdash In summary, by harnessing chaotic scrambling dynamics to correlate solid-state spins, we have engineered many-body correlated states that not only exhibit an exponentially growing QFI over the evolution time, but also demonstrate an exponentially amplified sensitivity to a global rotation, characterized by the same Lyapunov exponent. Our findings confirm the substantial metrological gains afforded by quantum scrambling and chaotic dynamics. Crucially, the scramblon-theory-guided analysis quantifies the competition between intrinsic correlation enhancement and the loss due to imperfect time reversal. This technique therefore enables characterization of metrological performance from generic chaotic scrambling dynamics across diverse experimental platforms such as cold atoms and nitrogen vacancy centers, where time-reversal imperfections are inevitably present.
	
	\ 
	
	\ 
	
\textit{Acknowledgements}---We thank Hui Zhai, Gonzalo A. Álvarez and Analia Zwick for helpful discussions. This work is supported by Quantum Science and Technology-National Science and Technology Major Project No.2021ZD0303205 (X.P.), National Natural Science Foundation of China No.12261160569 (X.P.), the XPLORER Prize (X.P.), and Fundamental Research Funds for the Central Universities under Grant No.WK2030250121 (Y.L.). This work was partially carried out at Instruments Center for Physical Science, University of Science and Technology of China.
\vspace{1em}
		
\textit{Data Availability}---The data that support the findings of this article are openly available \cite{code_zenodo}.

\bibliography{ref}% Produces the bibliography via 

@misc{SM,
note = {See Supplemental Material for further details regarding the experimental setup, the relationship between QFI and MQC, the Gaussian fitting of the MQC spectrum, and the phase-sensitivity calibration for the uncorrelated state.}
}

@dataset{code_zenodo,
  author= {Li, Yu-Chen and Zhang, Shengyu and Wu, Ze and Yin, Haochuan and Zhao, Liqiang and An, Xiaoxue and Cui, Jiaxi and Suter, Dieter and Peng, Xinhua},
  title        = {Data for ``{C}orrelation-enhanced metrology from scrambling dynamics in a solid-state spin system", {Z}enodo
                 },
  year         = 2026,
  publisher    = {Zenodo},
  doi={10.5281/zenodo.20808500},
}

@article{hardle1991bootstrap,
  title={Bootstrap simultaneous error bars for nonparametric regression},
  author={H{\"a}rdle, Wolfgang and Marron, James Stephen},
  journal={The Annals of Statistics},
  pages={778--796},
  year={1991},
  publisher={JSTOR},
  doi={10.1214/aos/1176348120}
}

@article{hauke2016measuring,
  title={Measuring multipartite entanglement through dynamic susceptibilities},
  author={Hauke, Philipp and Heyl, Markus and Tagliacozzo, Luca and Zoller, Peter},
  journal={Nat. Phys.},
  doi = {10.1038/nphys3700},
  volume={12},
  number={8},
  pages={778--782},
  year={2016},
  publisher={Nature Publishing Group UK London}
}

@book{abragam1961,
  author = {Abragam, Anatole},
  publisher = {Oxford university press},
  url = {https://global.oup.com/academic/product/the-principles-of-nuclear-magnetism-9780198520146?cc=jp&lang=en&},
  title = {The principles of nuclear magnetism},
  year = {1961}
}

@article{alvarez2015localization,
  author = {{\'A}lvarez, Gonzalo A and Suter, Dieter and Kaiser, Robin},
  doi = {10.1126/science.1261160},
  journal = {Science},
  number = {6250},
  pages = {846--848},
  publisher = {American Association for the Advancement of Science},
  title = {Localization-delocalization transition in the dynamics of dipolar-coupled nuclear spins},
  volume = {349},
  year = {2015}
}

@article{baum1985multiple,
  author = {Baum, J. and Munowitz, M. and Garroway, A. N. and Pines, A.},
  doi = {10.1063/1.449344},
  issn = {0021-9606},
  journal = {J. Chem. Phys.},
  month = {September},
  number = {5},
  pages = {2015--2025},
  publisher = {AIP Publishing},
  title = {Multiple-quantum Dynamics in Solid State {{NMR}}},
  urldate = {2025-06-17},
  volume = {83},
  year = {1985}
}

@article{blanes2009magnus,
  author = {Blanes, Sergio and Casas, Fernando and Oteo, Jose-Angel and Ros, Jos{\'e}},
  doi = {10.1016/j.physrep.2008.11.001},
  journal = {Phys. Rep.},
  number = {5-6},
  pages = {151--238},
  publisher = {Elsevier},
  title = {The Magnus expansion and some of its applications},
  volume = {470},
  year = {2009}
}

@article{braunstein1994statistical,
  author = {Braunstein, Samuel L and Caves, Carlton M},
  doi = {10.1103/PhysRevLett.72.3439},
  journal = {Phys. Rev. Lett.},
  number = {22},
  pages = {3439},
  publisher = {APS},
  title = {Statistical distance and the geometry of quantum states},
  volume = {72},
  year = {1994}
}

@article{CappellaroEmergentPrethermalizationSignatures2019,
  archiveprefix = {arXiv},
  author = {Wei, Ken Xuan and Peng, Pai and Shtanko, Oles and Marvian, Iman and Lloyd, Seth and Ramanathan, Chandrasekhar and Cappellaro, Paola},
  doi = {10.1103/PhysRevLett.123.090605},
  issn = {0031-9007, 1079-7114},
  journal = {Phys. Rev. Lett.},
  number = {9},
  pages = {090605},
  title = {Emergent Prethermalization Signatures in Out-of-Time Ordered Correlations},
  urldate = {2023-09-01},
  volume = {123},
  year = {2019}
}

@article{CappellaroExploringLocalizationNuclear2018a,
  author = {Wei, Ken Xuan and Ramanathan, Chandrasekhar and Cappellaro, Paola},
  doi = {10.1103/PhysRevLett.120.070501},
  journal = {Phys. Rev. Lett.},
  number = {7},
  pages = {070501},
  publisher = {American Physical Society},
  title = {Exploring {{Localization}} in {{Nuclear Spin Chains}}},
  urldate = {2025-05-30},
  volume = {120},
  year = {2018}
}

@article{chang1960heat,
  author = {Chang, Shu-Sing and Westrum Jr., Edgar F},
  doi = {10.1021/j100839a050},
  journal = {J. Chem. Phys.},
  pages = {1547--1551},
  publisher = {ACS Publications},
  title = {HEAT CAPACITIES AND THERMODYNAMIC PROPERTIES OF GLOBULAR MOLECULES. {I.} ADAMANTANE AND HEXAMETHYLENETETRAMINE},
  url = {https://doi.org/10.1021/j100839a050},
  volume = {64},
  year = {1960}
}

@article{ChenInformationScramblingQuantum2021,
  author = {Mi, Xiao and Roushan, Pedram and Quintana, Chris and \textit{et al.}},
  doi = {10.1126/science.abg5029},
  journal = {Science},
  number = {6574},
  pages = {1479--1483},
  publisher = {American Association for the Advancement of Science},
  title = {Information Scrambling in Quantum Circuits},
  urldate = {2025-05-30},
  volume = {374},
  year = {2021}
}

@article{ChinQuantumSimulationUnruh2019,
  author = {Hu, Jiazhong and Feng, Lei and Zhang, Zhendong and Chin, Cheng},
  copyright = {2019 The Author(s), under exclusive licence to Springer Nature Limited},
  doi = {10.1038/s41567-019-0537-1},
  issn = {1745-2481},
  journal = {Nat. Phys.},
  number = {8},
  pages = {785--789},
  publisher = {Nature Publishing Group},
  title = {Quantum Simulation of {{Unruh}} Radiation},
  urldate = {2025-05-31},
  volume = {15},
  year = {2019}
}

@article{DeutschQuantumStatisticalMechanics1991,
  author = {Deutsch, J. M.},
  doi = {10.1103/PhysRevA.43.2046},
  journal = {Phys. Rev. A},
  number = {4},
  pages = {2046--2049},
  publisher = {American Physical Society},
  title = {Quantum Statistical Mechanics in a Closed System},
  urldate = {2025-06-06},
  volume = {43},
  year = {1991}
}

@article{dominguez2021decoherence,
  author = {Dom\'{\i}nguez, Federico D. and Rodr\'{\i}guez, Mar\'{\i}a Cristina and Kaiser, Robin and Suter, Dieter and \'Alvarez, Gonzalo A.},
  doi = {10.1103/PhysRevA.104.012402},
  issue = {1},
  journal = {Phys. Rev. A},
  month = {Jul},
  numpages = {10},
  pages = {012402},
  publisher = {American Physical Society},
  title = {Decoherence scaling transition in the dynamics of quantum information scrambling},
  volume = {104},
  year = {2021}
}

@article{DuanInformationScramblingDynamics2022,
  author = {Wang, J.-H. and Cai, T.-Q. and Han, X.-Y. and Ma, Y.-W and Wang, Z.-L and Bao, Z.-H and Li, Y. and Wang, H.-Y and Zhang, H.-Y and Sun, L.-Y and Wu, Y.-K. and Song, Y.-P. and Duan, L.-M.},
  doi = {10.1103/PhysRevResearch.4.043141},
  journal = {Phys. Rev. Research},
  number = {4},
  pages = {043141},
  publisher = {American Physical Society},
  title = {Information Scrambling Dynamics in a Fully Controllable Quantum Simulator},
  urldate = {2025-05-30},
  volume = {4},
  year = {2022}
}

@article{DuEmergentUniversalQuench2024a,
  author = {Li, Yuchen and Zhou, Tian-Gang and Wu, Ze and Peng, Pai and Zhang, Shengyu and Fu, Riqiang and Zhang, Ren and Zheng, Wei and Zhang, Pengfei and Zhai, Hui and Peng, Xinhua and Du, Jiangfeng},
  copyright = {2024 The Author(s), under exclusive licence to Springer Nature Limited},
  doi = {10.1038/s41567-024-02664-0},
  issn = {1745-2481},
  journal = {Nat. Phys.},
  number = {12},
  pages = {1966--1972},
  publisher = {Nature Publishing Group},
  title = {Emergent Universal Quench Dynamics in Randomly Interacting Spin Models},
  urldate = {2025-03-15},
  volume = {20},
  year = {2024}
}

@book{duer2008solid,
  author = {Duer, Melinda J},
  doi = {10.1002/9780470999394},
  publisher = {John Wiley \& Sons},
  title = {Solid state NMR spectroscopy: principles and applications},
  url = {https://doi.org/10.1002/9780470999394},
  year = {2008}
}

@article{DuMeasuringOutofTimeOrderCorrelators2017a,
  author = {Li, Jun and Fan, Ruihua and Wang, Hengyan and Ye, Bingtian and Zeng, Bei and Zhai, Hui and Peng, Xinhua and Du, Jiangfeng},
  doi = {10.1103/PhysRevX.7.031011},
  journal = {Phys. Rev. X},
  number = {3},
  pages = {031011},
  publisher = {American Physical Society},
  title = {Measuring Out-of-Time-Order Correlators on a Nuclear Magnetic Resonance Quantum Simulator},
  urldate = {2025-06-06},
  volume = {7},
  year = {2017}
}

@article{gao2025signal,
  author = {Gao, Haoyang and Martin, Leigh S. and Hughes, Lillian B. and Leitao, Nathaniel T. and Put, Piotr and Zhou, Hengyun and Koyluoglu, Nazli U. and Meynell, Simon A. and Jayich, Ania C. Bleszynski and Park, Hongkun and Lukin, Mikhail D.},
  doi = {10.1038/s41586-025-09452-7},
  journal = {Nature},
  pages = {68--73},
  title = {Signal Amplification in a Solid-State Sensor through Asymmetric Many-Body Echo},
  url = {https://doi.org/10.1038/s41586-025-09452-7},
  volume = {646},
  year = {2025}
}

@article{garttner2018relating,
  author = {G{\"a}rttner, Martin and Hauke, Philipp and Rey, Ana Maria},
  doi = {10.1103/PhysRevLett.120.040402},
  journal = {Phys. Rev. Lett.},
  number = {4},
  pages = {040402},
  publisher = {APS},
  title = {Relating out-of-time-order correlations to entanglement via multiple-quantum coherences},
  volume = {120},
  year = {2018}
}

@misc{ge2025information,
  author = {Ge, Yangyang and Zhou, Haoyu and Zheng, Wen and Yu, Xiang-Min and Fang, Wei and Zhang, Zhenchuan and Huang, Wanli and Deng, Xiang and Cai, Haoyang and Li, Xianke and Zhou, Kun and Che, Hanxin and Zhang, Tao and Ji, Lichang and Zhang, Yu and Zhao, Jie and Li, Shao-Xiong and Tan, Xinsheng and Yu, Yang},
  archivePrefix = {arXiv},
  eprint = {2512.21157},
  title = {Information-Scrambling-Enhanced Quantum Sensing Beyond the Standard Quantum Limit},
}

@article{gorin2006dynamics,
  author = {Gorin, Thomas and Prosen, Toma{\v{z}} and Seligman, Thomas H and {\v{Z}}nidari{\v{c}}, Marko},
  doi = {10.1016/j.physrep.2006.09.003},
  journal = {Physics Reports},
  number = {2-5},
  pages = {33--156},
  publisher = {Elsevier},
  title = {Dynamics of Loschmidt echoes and fidelity decay},
  volume = {435},
  year = {2006}
}

@article{gu2019relation,
  author = {Gu, Yingfei and Kitaev, Alexei},
  doi = {10.1007/jhep02(2019)075},
  journal = {J. High Energy Phys.},
  pages = {75},
  publisher = {Springer},
  title = {On the relation between the magnitude and exponent of OTOCs},
  volume = {2019},
  year = {2019}
}

@article{gu2022two,
  author = {Gu, Yingfei and Kitaev, Alexei and Zhang, Pengfei},
  doi = {10.1007/JHEP03(2022)133},
  journal = {J. High Energy Phys.},
  pages = {133},
  title = {{A two-way approach to out-of-time-order correlators}},
  volume = {2022},
  year = {2022}
}

@article{helstrom1969quantum,
  author = {Helstrom, Carl W},
  doi = {10.1016/s0076-5392(08)x6017-5},
  journal = {Journal of Statistical Physics},
  number = {2},
  pages = {231--252},
  publisher = {Springer},
  title = {Quantum detection and estimation theory},
  volume = {1},
  year = {1969}
}

@article{hu2026quantum,
  author = {Hu, Guantian and Zhang, Wenxuan and Chen, Zhihua and Zhong, Liuzhu and Zhao, Jingchao and Liu, Chilong and Liu, Zixing and Xu, Yue and Lin, Yongchang and Ri, Yougui and Xie, Guixu and Liu, Mingze and Yuan, Haolan and Zhou, Yuxuan and Zhang, Yu and Hu, Chang-Kang and Liu, Song and Tan, Dian and Yu, Dapeng},
  doi = {10.1103/wb5z-y5y5},
  journal = {Phys. Rev. Lett.},
  number = {21},
  pages = {210801},
  publisher = {APS},
  title = {Quantum-Enhanced Sensing Enabled by Scrambling-Induced Genuine Multipartite Entanglement},
  volume = {136},
  year = {2026}
}

@article{jalabert2001environment,
  author = {Jalabert, Rodolfo A and Pastawski, Horacio M},
  doi = {10.1103/PhysRevLett.86.2490},
  journal = {Phys. Rev. Lett.},
  number = {12},
  pages = {2490},
  publisher = {APS},
  title = {Environment-independent decoherence rate in classically chaotic systems},
  volume = {86},
  year = {2001}
}

@misc{jyoti2017dipolarly,
  author = {Jyoti, Dhrubo},
  archiveprefix = {arXiv},
  eprint ={1711.01948},
  title = {Dipolarly-Coupled Chaotic Quantum Spin Systems},
}

@article{KaiserLocalizationdelocalizationTransitionDynamics2015,
  author = {{\'A}lvarez, Gonzalo A. and Suter, Dieter and Kaiser, Robin},
  doi = {10.1126/science.1261160},
  journal = {Science},
  number = {6250},
  pages = {846--848},
  publisher = {American Association for the Advancement of Science},
  title = {Localization-Delocalization Transition in the Dynamics of Dipolar-Coupled Nuclear Spins},
  urldate = {2025-05-30},
  volume = {349},
  year = {2015}
}

@article{Kaufman2016QuantumTT,
  author = {Adam M. Kaufman and M. Eric Tai and Alexander Lukin and Matthew Rispoli and R. Schittko and Philipp M. Preiss and Markus Greiner},
  doi = {10.1126/science.aaf6725},
  journal = {Science},
  pages = {794 - 800},
  title = {Quantum thermalization through entanglement in an isolated many-body system},
  url = {https://api.semanticscholar.org/CorpusID:206649361},
  volume = {353},
  year = {2016}
}

@inproceedings{kitaev2014hidden,
  note = {A. Kitaev, Hidden correlations in the Hawking radiation and thermal noise, Talk given at the Fundamental Physics Prize Symposium, 2014, \url{https://online.kitp.ucsb.edu/online/joint98/kitaev/}.}
}

@article{kitaev2018soft,
  author = {Kitaev, Alexei and Suh, S. Josephine},
  doi = {10.1007/JHEP05(2018)183},
  journal = {J. High Energy Phys.},
  pages = {183},
  publisher = {APS},
  title = {{The soft mode in the Sachdev-Ye-Kitaev model and its gravity dual}},
  volume = {2018},
}

@misc{kobrin2024universal,
  archivePrefix = {arXiv},
  author = {Kobrin, Bryce and Schuster, Thomas and Block, Maxwell and Wu, Weijie and Mitchell, Bradley and Davis, Emily and Yao, Norman Y},
  eprint = {2411.12794},
  title = {A universal protocol for quantum-enhanced sensing via information scrambling},
}

@article{kuwahara2016floquet,
  author = {Kuwahara, Tomotaka and Mori, Takashi and Saito, Keiji},
  doi = {10.1016/j.aop.2016.01.012},
  journal = {Annals of Physics},
  pages = {96--124},
  publisher = {Elsevier},
  title = {Floquet--Magnus theory and generic transient dynamics in periodically driven many-body quantum systems},
  volume = {367},
  year = {2016}
}

@article{larkin1969quasiclassical,
  author = {Larkin, AI and Ovchinnikov, Yu N},
  journal = {Sov Phys JETP},
  number = {6},
  pages = {1200--1205},
  title = {Quasiclassical method in the theory of superconductivity},
  url = {http://jetp.ras.ru/cgi-bin/dn/e_028_06_1200.pdf},
  volume = {28},
  year = {1969}
}

@article{li2026errorresilient,
  author = {Li, Yu-Chen and Zhou, Tian-Gang and Zhang, Shengyu and Wu, Ze and Zhao, Liqiang and Yin, Haochuan and An, Xiaoxue and Zhai, Hui and Zhang, Pengfei and Peng, Xinhua and Du, Jiangfeng}, 
  doi = {10.1103/cg3f-rggs},
  journal = {Phys. Rev. Lett.},
  number = {6},
  pages = {060403},
  publisher = {APS},
  title = {Error-resilient reversal of quantum chaotic dynamics enabled by scramblons},
  volume = {136},
  year = {2026}
}

@article{LiExperimentalObservationEquilibrium2020,
  author = {Nie, Xinfang and Wei, Bo-Bo and Chen, Xi and Zhang, Ze and Zhao, Xiuzhu and Qiu, Chudan and Tian, Yu and Ji, Yunlan and Xin, Tao and Lu, Dawei and Li, Jun},
  doi = {10.1103/PhysRevLett.124.250601},
  journal = {Phys. Rev. Lett.},
  number = {25},
  pages = {250601},
  publisher = {American Physical Society},
  title = {Experimental Observation of Equilibrium and Dynamical Quantum Phase Transitions via Out-of-Time-Ordered Correlators},
  urldate = {2025-05-29},
  volume = {124},
  year = {2020}
}

@article{LinkeExperimentalMeasurementOutofTimeOrdered2022,
  author = {Green, Alaina M. and Elben, A. and Alderete, C. Huerta and Joshi, Lata Kh and Nguyen, Nhung H. and Zache, Torsten V. and Zhu, Yingyue and Sundar, Bhuvanesh and Linke, Norbert M.},
  doi = {10.1103/PhysRevLett.128.140601},
  journal = {Phys. Rev. Lett.},
  number = {14},
  pages = {140601},
  publisher = {American Physical Society},
  title = {Experimental Measurement of Out-of-Time-Ordered Correlators at Finite Temperature},
  urldate = {2025-05-29},
  volume = {128},
  year = {2022}
}

@misc{LiObservationQuantumInformation2024a,
  archiveprefix = {arXiv},
  author = {Xiang, De-Sheng and Zhang, Yao-Wen and Liu, Hao-Xiang and Zhou, Peng and Yuan, Dong and Zhang, Kuan and Zhang, Shun-Yao and Xu, Biao and Liu, Lu and Li, Yitong and Li, Lin},
  eprint = {2410.15455},
  number = {arXiv:2410.15455},
  publisher = {arXiv},
  title = {Observation of Quantum Information Collapse-and-Revival in a Strongly-Interacting {{Rydberg}} Atom Array},
  urldate = {2025-05-29},
}

@article{Liu2023sig,
  author = {Liu, Zeyu and Zhang, Pengfei},
  doi = {10.1103/PhysRevLett.132.060201},
  issn = {0031-9007, 1079-7114},
  journal = {Phys. Rev. Lett.},
  number = {6},
  pages = {060201},
  title = {Signature of {{Scramblon Effective Field Theory}} in {{Random Spin Models}}},
  volume = {132},
  year = {2024}
}

@article{liu2025variational,
  author = {Liu, Ran and Wu, Ze and Yang, Xiaodong and Li, Yuchen and Zhou, Hui and Li, Zhaokai and Chen, Yuquan and Yuan, Haidong and Peng, Xinhua},
  copyright = {https://creativecommons.org/licenses/by/4.0/},
  doi = {10.1093/nsr/nwaf091},
  issn = {2095-5138, 2053-714X},
  journal = {Natl. Sci. Rev.},
  number = {5},
  pages = {nwaf091},
  title = {Variational Quantum Metrology with the {{Loschmidt}} Echo},
  volume = {12},
  year = {2025}
}

@article{macri2016loschmidt,
  author = {Macr{\`\i}, Tommaso and Smerzi, Augusto and Pezz{\`e}, Luca},
  doi = {10.1103/PhysRevA.94.010102},
  journal = {Phys. Rev. A},
  number = {1},
  pages = {010102},
  publisher = {APS},
  title = {Loschmidt echo for quantum metrology},
  volume = {94},
  year = {2016}
}

@article{Maldacena_2016,
  author = {Maldacena, Juan and Shenker, Stephen H. and Stanford, Douglas},
  doi = {10.1007/jhep08(2016)106},
  issn = {1029-8479},
  journal = {J. High Energ. Phys.},
  month = {August},
  pages = {106},
  publisher = {Springer Science and Business Media LLC},
  title = {A bound on chaos},
  url = {http://dx.doi.org/10.1007/JHEP08(2016)106},
  volume = {2016},
  year = {2016}
}

@article{mccall1960nuclear,
  author = {McCall, David W and Douglass, Dean C},
  doi = {10.1063/1.1731259},
  journal = {J. Chem. Phys.},
  pages = {777--778},
  title = {Nuclear magnetic resonance in solid adamantane},
  url = {https://doi.org/10.1063/1.1731259},
  volume = {33},
  year = {1960}
}

@article{modi2011quantum,
  author = {Modi, Kavan and Cable, Hugo and Williamson, Mark and Vedral, Vlatko},
  doi = {10.1103/PhysRevX.1.021022},
  journal = {Physical Review X},
  number = {2},
  pages = {021022},
  publisher = {APS},
  title = {Quantum correlations in mixed-state metrology},
  volume = {1},
  year = {2011}
}

@article{MonroeVerifiedQuantumInformation2019,
  author = {Landsman, K. A. and Figgatt, C. and Schuster, T. and Linke, N. M. and Yoshida, B. and Yao, N. Y. and Monroe, C.},
  copyright = {2019 The Author(s), under exclusive licence to Springer Nature Limited},
  doi = {10.1038/s41586-019-0952-6},
  issn = {1476-4687},
  journal = {Nature},
  number = {7746},
  pages = {61--65},
  publisher = {Nature Publishing Group},
  title = {Verified Quantum Information Scrambling},
  urldate = {2025-05-29},
  volume = {567},
  year = {2019}
}

@article{nordman1965phase,
  author = {Nordman, CE and Schmitkons, DL},
  doi = {10.1107/s0365110x65001755},
  journal = {Acta Crystallogr.},
  pages = {764--767},
  publisher = {International Union of Crystallography},
  title = {Phase transition and crystal structures of adamantane},
  url = {https://doi.org/10.1107/S0365110X65001755},
  volume = {18},
  year = {1965}
}

@article{OliverProbingQuantumInformation2022,
  author = {Braum{\"u}ller, Jochen and Karamlou, Amir H. and Yanay, Yariv and Kannan, Bharath and Kim, David and Kjaergaard, Morten and Melville, Alexander and Niedzielski, Bethany M. and Sung, Youngkyu and Veps{\"a}l{\"a}inen, Antti and Winik, Roni and Yoder, Jonilyn L. and Orlando, Terry P. and Gustavsson, Simon and Tahan, Charles and Oliver, William D.},
  copyright = {2021 The Author(s), under exclusive licence to Springer Nature Limited},
  doi = {10.1038/s41567-021-01430-w},
  issn = {1745-2481},
  journal = {Nat. Phys.},
  number = {2},
  pages = {172--178},
  publisher = {Nature Publishing Group},
  title = {Probing Quantum Information Propagation with Out-of-Time-Ordered Correlators},
  urldate = {2025-05-29},
  volume = {18},
  year = {2022}
}

@article{PastawskiEmergentDecoherenceInduced2022,
  author = {S{\'a}nchez, C. M. and Chattah, A. K. and Pastawski, H. M.},
  doi = {10.1103/PhysRevA.105.052232},
  journal = {Phys. Rev. A},
  number = {5},
  pages = {052232},
  publisher = {American Physical Society},
  shorttitle = {Emergent Decoherence Induced by Quantum Chaos in a Many-Body System},
  title = {Emergent Decoherence Induced by Quantum Chaos in a Many-Body System: {{A Loschmidt}} Echo Observation through {{NMR}}},
  urldate = {2025-05-30},
  volume = {105},
  year = {2022}
}

@article{resing1969nmr,
  author = {Resing, HA},
  doi = {10.1080/15421406908082735},
  journal = {Mol. Cryst. Liq. Cryst.},
  number = {1},
  pages = {101--132},
  publisher = {Taylor \& Francis},
  title = {{NMR} relaxation in adamantane and hexamethylenetetramine: diffusion and rotation},
  url = {https://doi.org/10.1080/15421406908082735},
  volume = {9},
  year = {1969}
}

@article{ReyMeasuringOutoftimeorderCorrelations2017a,
  author = {G{\"a}rttner, Martin and Bohnet, Justin G. and {Safavi-Naini}, Arghavan and Wall, Michael L. and Bollinger, John J. and Rey, Ana Maria},
  copyright = {2017 Springer Nature Limited},
  doi = {10.1038/nphys4119},
  issn = {1745-2481},
  journal = {Nat. Phys.},
  number = {8},
  pages = {781--786},
  publisher = {Nature Publishing Group},
  title = {Measuring Out-of-Time-Order Correlations and Multiple Quantum Spectra in a Trapped-Ion Quantum Magnet},
  urldate = {2025-05-29},
  volume = {13},
  year = {2017}
}

@article{ReyUnifyingScramblingThermalization2019b,
  author = {{Lewis-Swan}, R. J. and {Safavi-Naini}, A. and Bollinger, J. J. and Rey, A. M.},
  copyright = {2019 This is a U.S. government work and not under copyright protection in the U.S.; foreign copyright protection may apply},
  doi = {10.1038/s41467-019-09436-y},
  issn = {2041-1723},
  journal = {Nat. Commun.},
  number = {1},
  pages = {1581},
  publisher = {Nature Publishing Group},
  title = {Unifying Scrambling, Thermalization and Entanglement through Measurement of Fidelity out-of-Time-Order Correlators in the {{Dicke}} Model},
  urldate = {2025-05-30},
  volume = {10},
  year = {2019}
}

@article{rhim1970violation,
  author = {Rhim, W-K and Pines, Alexander and Waugh, John S},
  doi = {10.1103/PhysRevLett.25.218},
  journal = {Phys. Rev. Lett.},
  number = {4},
  pages = {218},
  publisher = {APS},
  title = {Violation of the spin-temperature hypothesis},
  url = {https://doi.org/10.1103/PhysRevLett.25.218},
  volume = {25},
  year = {1970}
}

@article{Rigol2007ThermalizationAI,
  author = {Marcos Rigol and Vanja Dunjko and Maxim Olshanii},
  doi = {10.1038/nature06838},
  journal = {Nature},
  pages = {854-858},
  title = {Thermalization and its mechanism for generic isolated quantum systems},
  volume = {452},
  year = {2007}
}

@article{robertsLocalizedShocks2015,
  author = {Roberts, Daniel A. and Stanford, Douglas and Susskind, Leonard},
  doi = {10.1007/JHEP03(2015)051},
  issn = {1029-8479},
  journal = {J. High Energ. Phys.},
  pages = {51},
  title = {Localized Shocks},
  volume = {2015},
  year = {2015}
}

@article{sanchez2020perturbation,
  author = {S\'anchez, C. M. and Chattah, A. K. and Wei, K. X. and Buljubasich, L. and Cappellaro, P. and Pastawski, H. M.},
  doi = {10.1103/PhysRevLett.124.030601},
  issue = {3},
  journal = {Phys. Rev. Lett.},
  month = {Jan},
  numpages = {6},
  pages = {030601},
  publisher = {American Physical Society},
  title = {Perturbation Independent Decay of the Loschmidt Echo in a Many-Body System},
  volume = {124},
  year = {2020}
}

@misc{SDBSWeb,
  author = {},
  title = {SDBSWeb : https://sdbs.db.aist.go.jp (National Institute of Advanced Industrial Science and Technology)},
  url = {https://sdbs.db.aist.go.jp},
  year = {2025}
}

@article{shenkerBlackHolesButterfly2014,
  author = {Shenker, Stephen H. and Stanford, Douglas},
  doi = {10.1007/JHEP03(2014)067},
  issn = {1029-8479},
  journal = {J. High Energ. Phys.},
  pages = {67},
  title = {Black Holes and the Butterfly Effect},
  volume = {2014},
  year = {2014}
}

@article{shenkerStringyEffectsScrambling2015,
  author = {Shenker, Stephen H. and Stanford, Douglas},
  doi = {10.1007/JHEP05(2015)132},
  issn = {1029-8479},
  journal = {J. High Energ. Phys.},
  pages = {132},
  title = {Stringy Effects in Scrambling},
  volume = {2015},
  year = {2015}
}

@article{smith1961calculation,
  author = {Smith, George W},
  doi = {10.1063/1.1701195},
  journal = {J. Chem. Phys.},
  number = {3},
  pages = {1134--1135},
  publisher = {American Institute of Physics},
  title = {On the calculation of second moments of nuclear magnetic resonance lines for large molecules. Adamantane molecule},
  url = {https://doi.org/10.1063/1.1701195},
  volume = {35},
  year = {1961}
}

@article{SrednickiApproachThermalEquilibrium1999,
  archiveprefix = {arXiv},
  author = {Srednicki, Mark},
  doi = {10.1088/0305-4470/32/7/007},
  issn = {0305-4470, 1361-6447},
  journal = {J. Phys. A: Math. Gen.},
  number = {7},
  pages = {1163--1175},
  title = {The Approach to Thermal Equilibrium in Quantized Chaotic Systems},
  urldate = {2025-06-06},
  volume = {32},
  year = {1999}
}

@article{stanford2022subleading,
  author = {Stanford, Douglas and Yang, Zhenbin and Yao, Shunyu},
  doi = {10.1007/jhep02(2022)200},
  journal = {J. High Energy Phys.},
  pages = {200},
  publisher = {Springer},
  title = {Subleading weingartens},
  volume = {2022},
  year = {2022}
}

@article{Swingle:2018ekw,
  author = {Swingle, Brian},
  doi = {10.1038/s41567-018-0295-5},
  journal = {Nat. Phys.},
  number = {10},
  pages = {988--990},
  title = {{Unscrambling the physics of out-of-time-order correlators}},
  volume = {14},
  year = {2018}
}

@article{ThomasEnergyResolvedInformationScrambling2021,
  author = {Pegahan, S. and Arakelyan, I. and Thomas, J. E.},
  doi = {10.1103/PhysRevLett.126.070601},
  journal = {Phys. Rev. Lett.},
  number = {7},
  pages = {070601},
  publisher = {American Physical Society},
  title = {Energy-Resolved Information Scrambling in Energy-Space Lattices},
  urldate = {2025-05-29},
  volume = {126},
  year = {2021}
}

@article{VuleticImprovingMetrologyQuantum2023,
  author = {Li, Zeyang and Colombo, Simone and Shu, Chi and Velez, Gustavo and {Pilatowsky-Cameo}, Sa{\'u}l and Schmied, Roman and Choi, Soonwon and Lukin, Mikhail and {Pedrozo-Pe{\~n}afiel}, Edwin and Vuleti{\'c}, Vladan},
  doi = {10.1126/science.adg9500},
  journal = {Science},
  number = {6652},
  pages = {1381--1384},
  publisher = {American Association for the Advancement of Science},
  title = {Improving Metrology with Quantum Scrambling},
  urldate = {2023-11-14},
  volume = {380},
  year = {2023}
}

@article{VuleticTimereversalbasedQuantumMetrology2022,
  author = {Colombo, Simone and {Pedrozo-Pe{\~n}afiel}, Edwin and Adiyatullin, Albert F. and Li, Zeyang and Mendez, Enrique and Shu, Chi and Vuleti{\'c}, Vladan},
  copyright = {2022 The Author(s), under exclusive licence to Springer Nature Limited},
  doi = {10.1038/s41567-022-01653-5},
  issn = {1745-2481},
  journal = {Nat. Phys.},
  number = {8},
  pages = {925--930},
  publisher = {Nature Publishing Group},
  title = {Time-Reversal-Based Quantum Metrology with Many-Body Entangled States},
  urldate = {2023-12-11},
  volume = {18},
  year = {2022}
}

@article{wang2021triple,
  author = {Wang, Fenfen and Ramakrishna, Sanath K and Sun, Pingchuan and Fu, Riqiang},
  doi = {10.1016/j.jmr.2021.107067},
  journal = {J. Magn. Reson.},
  pages = {107067},
  publisher = {Elsevier},
  title = {Triple-pulse excitation: An efficient way for suppressing background signals and eliminating radio-frequency acoustic ringing in direct polarization {NMR} experiments},
  url = {https://doi.org/10.1016/j.jmr.2021.107067},
  volume = {332},
  year = {2021}
}

@article{WeidemullerTimereversalDipolarQuantum2024,
  author = {Geier, Sebastian and Braemer, Adrian and Braun, Eduard and M{\"u}llenbach, Maximilian and Franz, Titus and G{\"a}rttner, Martin and Z{\"u}rn, Gerhard and Weidem{\"u}ller, Matthias},
  doi = {10.1103/PhysRevResearch.6.033197},
  journal = {Phys. Rev. Research},
  number = {3},
  pages = {033197},
  publisher = {American Physical Society},
  title = {Time-Reversal in a Dipolar Quantum Many-Body Spin System},
  urldate = {2025-05-28},
  volume = {6},
  year = {2024}
}

@article{yan2020information,
  author = {Yan, Bin and Cincio, Lukasz and Zurek, Wojciech H},
  doi = {10.1103/PhysRevLett.124.160603},
  journal = {Phys. Rev. Lett.},
  number = {16},
  pages = {160603},
  publisher = {APS},
  title = {Information scrambling and loschmidt echo},
  volume = {124},
  year = {2020}
}

@article{yen1983multiple,
  author = {Yen, Yu-Sze and Pines, A.},
  doi = {10.1063/1.445185},
  issn = {0021-9606},
  journal = {J. Chem. Phys.},
  month = {March},
  number = {6},
  pages = {3579--3582},
  publisher = {AIP Publishing},
  title = {Multiple-quantum {{NMR}} in Solids},
  urldate = {2025-06-17},
  volume = {78},
  year = {1983}
}

@article{YouObservationAnomalousInformation2024a,
  author = {Liang, Xinhui and Yue, Zongpei and Chao, Yu-Xin and Hua, Zhen-Xing and Lin, Yige and Tey, Meng Khoon and You, Li},
  doi = {10.1103/w1cp-l5vq},
  journal = {Phys. Rev. Lett.},
  number = {7},
  pages = {050201},
  publisher = {American Physical Society},
  title = {Observation of Anomalous Information Scrambling in a {{Rydberg}} Atom Array},
  url = {https://link.aps.org/doi/10.1103/w1cp-l5vq},
  volume = {135},
  year = {2025}
}

@article{ZhaoProbingOperatorSpreading2022,
  author = {Zhao, S. K. and Ge, Zi-Yong and Xiang, Zhongcheng and Xue, G. M. and Yan, H. S. and Wang, Z. T. and Wang, Zhan and Xu, H. K. and Su, F. F. and Yang, Z. H. and Zhang, He and Zhang, Yu-Ran and Guo, Xue-Yi and Xu, Kai and Tian, Ye and Yu, H. F. and Zheng, D. N. and Fan, Heng and Zhao, S. P.},
  doi = {10.1103/PhysRevLett.129.160602},
  journal = {Phys. Rev. Lett.},
  number = {16},
  pages = {160602},
  publisher = {American Physical Society},
  title = {Probing Operator Spreading via Floquet Engineering in a Superconducting Circuit},
  urldate = {2025-05-30},
  volume = {129},
  year = {2022}
}

@article{zhou2023operator,
  author = {Zhou, Tianci and Swingle, Brian},
  doi = {10.1038/s41467-023-39065-5},
  journal = {Nat. Commun.},
  number = {1},
  pages = {3411},
  publisher = {Nature Publishing Group UK London},
  title = {Operator growth from global out-of-time-order correlators},
  volume = {14},
  year = {2023}
}

@article{ZobristConstructiveInterferenceEdge2025,
  author = {{Google Quantum AI and Collaborators}},
  doi = {10.1038/s41586-025-09526-6},
  journal = {Nature},
  pages = {825--830},
  title = {Observation of Constructive Interference at the Edge of Quantum Ergodicity},
  url = {https://doi.org/10.1038/s41586-025-09526-6},
  volume = {646},
  year = {2025}
}

\newpage

\section{End matter}

\textit{Relation between QFI and MQC}---
For a general mixed probe state with eigendecomposition $\hat{\rho}=\sum_{k}p_k\ket{\psi_k}\bra{\psi_k}$, the QFI is defined as~\cite{braunstein1994statistical,helstrom1969quantum}
\begin{equation}
\mathcal{F}_{\rm Q}(\hat{\rho},\hat{O})\equiv 2\sum_{k,l}\frac{\left(p_k-p_l\right)^2}{p_k+p_l}|\bra{\psi_k}\hat{O}\ket{\psi_l}|^2,
\label{eq:QFI_raw}
\end{equation}
where $\hat{O}\equiv{\rm i}\hat{U}_\alpha^\dagger(\partial\hat{U}_\alpha/\partial \alpha)$ is the generator of the parameter-dependent unitary encoding $\hat{U}_\alpha$. In our specific protocol, the phase-shift encoding is $\hat{U}_\phi={\rm e}^{-{\rm i}\hat{O}_z\phi}$, making the collective spin operator $\hat{O}_z$ the relevant generator. 

The initial state is $\hat{\rho}_0=\mathrm{e}^{\epsilon\hat{O}_z}/\mathcal{Z} \approx (\mathbbm{1} + \epsilon \hat{O}_z)/2^N$, and the evolved probe state is $\hat{\rho}_t={\rm e}^{-{\rm i}\hat{H}t}\hat{\rho}_0{\rm e}^{{\rm i}\hat{H}t}$. Meanwhile, the MQC intensity of order $m$, expressed in the eigenbasis of $\hat{O}_z$, is given by
\begin{equation}
    I_m(t)=\frac{1}{\mathcal{C}}\sum_{\substack{k,l \text { s.t. } \\ M_k-M_l=m}}\left|\bra{\psi_k}(\hat{\rho}_t-\hat{\mathbbm{1}} / 2^N)\ket{\psi_l}\right|^2,
    \label{eq:MQC}
\end{equation}
where $\{M_k\}$ and $\{\ket{\psi_k}\}$ denote the eigenvalues and eigenvectors of $\hat{O}_z$, respectively, and $\mathcal{C}$ is a normalization constant ensuring $\sum_m I_m(t)=1$.

Because unitary evolution preserves the density matrix spectrum, the eigenvalues of $\hat{\rho}_t$ are identical to those of the initial state. We thus have $p_k(t)=p_k(0)=\mathrm{e}^{\epsilon M_k}/\mathcal{Z}$ and $\ket{\psi_k(t)}={\rm e}^{-{\rm i}\hat{H}t}\ket{\psi_k(0)}$. Substituting these relations into Eq.~\eqref{eq:QFI_raw}, we obtain 
\begin{equation}
    \mathcal{F}_{\rm Q}
    =2\sum_{k,l}\tanh^2\left[\frac{\epsilon(M_k-M_l)}{2}\right](p_k+p_l)\abs{\bra{\psi_k}\hat{O}_z(-t)\ket{\psi_l}}^2,
\end{equation}
which reproduces the result derived in Ref.~\onlinecite{hauke2016measuring} for evaluating the QFI of thermal probe states. 

In our system, the correlated spin clusters possess a finite size $K(t)$, which naturally bounds the support of the Heisenberg-evolved operator $\hat{O}_z(-t)$. Thus, the total $N$ spins can be effectively decomposed into $N/K$ disentangled $K$-spin clusters, yielding $\mathcal{F}_{\rm Q}=(N/K)\mathcal{F}_{\rm Q}^{(K)}$. Given the experimental parameters $\epsilon\sim 10^{-5}$ and $K(t)\lesssim 10^4$, the condition $\epsilon K\ll 1$ is satisfied. Therefore, within the Hilbert space of a single $K$-spin cluster, we can approximate $p_k+p_l\approx [2+\epsilon( M_k+M_l)]/2^K = [2+\mathrm{O}(K\epsilon)]/2^K$. This leads to
\begin{equation}
\begin{aligned}
\mathcal{F}_{\rm Q}^{(K)}=&\frac{4+\mathrm{O}(K\epsilon)}{2^K}\sum_{m=-K}^{K}\left(\tanh{\frac{\epsilon m}{2}}\right)^2\sum_{\substack{k,l \text { s.t. } \\ M_k-M_l=m}}|\bra{\psi_k}\hat{O}_z(-t)\ket{\psi_l}|^2\\
=&\frac{4+\mathrm{O}(K\epsilon)}{2^K}[\Tr_K(\hat{O}_z^2)]\sum_{m=-K}^{K} \left(\tanh{\frac{\epsilon m}{2}}\right)^2 I_m(t)\\
=&[1+\mathrm{O}(K\epsilon)] K\sum_{m=-K}^{K}\left(\tanh{\frac{\epsilon m}{2}}\right)^2I_m(t).
\end{aligned}
\end{equation}
These intermediate steps follow from the definition of the MQC intensity in Eq.~\eqref{eq:MQC}, the trace identity $\Tr_K(\hat{O}_z^2)=K2^K/4$, and the symmetry $I_m(t)=I_m(-t)$. Furthermore, since $\epsilon K\ll 1$, we can approximate $\tanh\left(\frac{\epsilon m}{2}\right)\approx\frac{\epsilon m}{2}$, which ultimately yields
\begin{equation}
\begin{aligned}
    \mathcal{F}_{\rm Q}(\hat{\rho}_t,\hat{O}_z)=& \frac{N}{K}\mathcal{F}_{\rm Q}^{(K)}=[1+\mathrm{O}(K\epsilon)]\frac{N}{4}\epsilon^2 \sum_m m^2 I_m(t).
    \end{aligned}
\end{equation}

\end{document}

\title{Supplementary Material}
\maketitle
\tableofcontents

\section{Experimental setup}

\subsection{System Hamiltonian}
Experiments were performed on a 400-MHz NMR spectrometer using a powdered adamantane (C\textsubscript{10}H\textsubscript{16}) sample, focusing on the network of spin-1/2 $^{1}$H nuclei. The most abundant carbon isotope, $^{12}$C at 99\% natural abundance, is spin-zero and NMR-silent. The sample was held at room temperature in a homogeneous $B_0=9.4$~T field aligned along the $z$ axis, produced by a superconducting magnet. The measured spin--lattice relaxation time is $T_1 = 1.3$~s.

At room temperature adamantane forms a face-centered cubic (fcc) plastic crystal (space group Fm3m)~\cite{nordman1965phase,mccall1960nuclear,chang1960heat}, in which the nearly spherical molecules reorient rapidly and isotropically on a picosecond timescale ($\sim 10^{-11}$~s)~\cite{resing1969nmr}. This rapid tumbling averages intra-molecular dipole--dipole couplings to zero, so the spin dynamics are governed by \emph{inter}-molecular couplings between protons on different molecules. To a good approximation the spins may be located at the molecular centers~\cite{resing1969nmr,mccall1960nuclear,smith1961calculation}.

In the laboratory frame the total $^{1}$H Hamiltonian reads
\begin{equation}
    \hat{H}_{\text{lab}}=\hat{H}_{\rm Z}+\hat{H}_{\mathrm{CS}}+\hat{H}_{\rm rf}(t)+\hat{H}_{\mathrm{dip}},
    \label{LabSysHam}
\end{equation}
where the four terms represent, respectively, the Zeeman interaction, chemical shifts and field inhomogeneities, the time-dependent RF pulses, and the dipolar interaction.

The dominant term is the Zeeman Hamiltonian $\hat{H}_{\rm Z} = \omega_{\rm H}\hat{O}_z$, with Larmor frequency $|\omega_{\rm H}|/(2\pi) = |\gamma_{\rm H}B_0|/(2\pi) = 400.15$~MHz ($\gamma_{\rm H}$ is the proton gyromagnetic ratio) and $\hat{O}_z = \sum_{i\mu}^N \hat{S}_{i\mu}^z$, where $\hat{S}_{i\mu}^\alpha$ ($\alpha=x,y,z$) denotes the spin operator for spin $\mu$ on molecule $i$, and $N$ is the total number of spins. Each powder granule (size of several micrometers) contains roughly $10^{10}$--$10^{13}$ proton spins~\cite{DuEmergentUniversalQuench2024a}. The second term, $\hat{H}_{\mathrm{CS}}$, describes small frequency offsets from chemical shifts and field inhomogeneities: the chemical-shift dispersion of adamantane is tens of Hertz~\cite{SDBSWeb} and inhomogeneous broadening contributes less than 12~Hz linewidth. These small shifts are effectively refocused by the multi-pulse sequences employed and are neglected in what follows.

The RF Hamiltonian induced by the pulses is
\begin{equation}
    \hat{H}_{\rm rf}(t)=\omega_{1}\bigl[\cos(\omega_{\rm rf}t+\alpha)\,\hat{O}_{x}+\sin(\omega_{\rm rf}t+\alpha)\,\hat{O}_{y}\bigr],
\end{equation}
with nutation frequency $\omega_1$ and pulse phase $\alpha$; we always operate on resonance, $\omega_{\rm rf}=\omega_{\rm H}$.

The fourth term is the inter-molecular dipole--dipole coupling:
\begin{equation}
\hat{H}_{\mathrm{dip}}=\sum_{i<j,\mu,\nu}\frac{\mu_0\hbar^2\gamma^2_\text{H}}{4\pi R^3_{ij}}
\Bigl[\hat{\bm{S}}_{i\mu}\!\cdot\!\hat{\bm{S}}_{j\nu}
-\frac{3(\hat{\bm{S}}_{i\mu}\!\cdot\!{\bm R}_{ij})(\hat{\bm{S}}_{j\nu}\!\cdot\!{\bm R}_{ij})}{R^2_{ij}}\Bigr],
\end{equation}
where $\mu_0$ is the vacuum permeability, $\gamma_\text{H}$ the proton gyromagnetic ratio, and $\bm{R}_{ij}$ the vector connecting the centers of the molecules, with $R_{ij}=|\bm{R}_{ij}|$.

In our experiment, $|\omega_{\rm H}| \gg \omega_1$ ($\omega_1\sim 0.1$ MHz) and is also much larger than the local dipolar field ($\sim$ kHz). It is therefore conventional to work in the interaction picture with respect to $\hat{H}_{\rm Z}$ (equivalent to a rotating-frame analysis) and retain only the components of $\hat{H}_{\mathrm{dip}}$ that commute with $\hat{O}_z$. For the radio-frequency pulse on resonance ($\omega_{\rm rf}=\omega_{\rm H}$), the dominant contribution is 
\begin{equation}
    \hat{H}_{\rm rf}^{\rm rot}=\omega_{1}[\hat{O}_{x}\cos\alpha+\hat{O}_{y}\sin\alpha].
\end{equation}
We apply discrete pulses of finite duration to induce collective rotations along the axis specified by $\alpha$, while during the intervals between pulses $\hat{H}_{\rm rf}^{\rm rot}=0$. The system's evolution during the free evolution between RF pulses is governed by the secular dipolar Hamiltonian~\cite{abragam1961,duer2008solid}
\begin{align}
    \hat{H}_{\mathrm{dd}} \equiv \sum_{i<j,\mu,\nu} J_{ij} (3\hat{S}^{z}_{i\mu}\hat{S}^{z}_{j\nu} - \hat{\bm{S}}_{i\mu}\cdot\hat{\bm{S}}_{j\nu}),
    \label{eq:Hdd}
\end{align}
where 
\begin{equation}
    J_{ij} = \left(\frac{\mu_0}{4\pi}\right)\frac{\hbar\gamma_{\mathrm{H}}^{2}}{2R_{ij}^3}\left(1-3\cos^2\theta_{ij}\right).
    \label{eq:def_Jij}
\end{equation}
$\theta_{ij}$ is the angle between $\bm{R}_{ij}$ and the external magnetic field direction $z$. The powder average over random orientations guarantees that $\overline{J_{ij}}=0$ since 
\begin{equation}
    \overline{1-3\cos^2\theta}=\int_0^{\pi}\dd \theta(1-3\cos^2\theta)\sin\theta =0.
\end{equation}
We define the effective dipolar coupling strength $J$ as
\begin{equation}
    J\equiv\sqrt{\frac{N\overline{J_{ij}^2}}{4}},
\end{equation}
where $J\approx 1460$~Hz has been carefully calibrated experimentally (see the supplementary material of Refs.~\onlinecite{DuEmergentUniversalQuench2024a,li2026errorresilient}). Throughout this work we use $Jt$ as the dimensionless time scale.

\subsection{Hamiltonian engineering}
To engineer the target Hamiltonians, we periodically apply RF pulse sequences to the system. The sequence we use is the 8-pulse sequence~\cite{yen1983multiple,baum1985multiple}

\begin{equation}
	\left(\frac{\tau}{2}, \mathbf{Y}, \tau', \mathbf{Y}, \tau, \mathbf{Y}, \tau', \mathbf{Y}, \tau, \overline{\mathbf{Y}}, \tau', \overline{\mathbf{Y}}, \tau, \overline{\mathbf{Y}}, \tau', \overline{\mathbf{Y}}, \frac{\tau}{2}\right),
    \label{eq:8pulse}
\end{equation}
where $\mathbf{Y}=\mathrm{e}^{-\mathrm{i}\hat{O}_y\pi/2}$ and $\overline{\mathbf{Y}}=\mathrm{e}^{\mathrm{i}\hat{O}_y\pi/2}$ denote $\pi/2$ pulses applied about $x$ or $y$ axis, with width $\tau_{\text{p}}=2\ \mu\text{s}$. The pulse interval $\tau'=2\tau+\tau_{\text{p}}$ with $\tau=2\ \mu{\rm s}$. During the intervals the system undergoes free evolution under the secular dipolar Hamiltonian $\hat{H}_{\rm dd}$ (Eq.~\eqref{eq:Hdd}). The pulses are applied from right to left in chronological order with a period of $T=12(\tau+\tau_{\rm p})=48~\mu s$. If the system is under stroboscopic observation at multiple periods, its dynamics will be governed by an effective average Hamiltonian which can be calculated via the Floquet-Magnus expansion \cite{kuwahara2016floquet, blanes2009magnus}, i.e., $\hat{H}=\sum_n\hat{H}_{\text{F}}^{(n)}$. Each term can be evaluated in a toggling frame that rotates in synchrony with the applied pulses \cite{duer2008solid}. In this frame, the dipolar Hamiltonian $\hat{H}_{\text{dd}}$ is transformed into a form that is piecewise-constant during the pulse intervals:
\begin{equation}
    \hat{H}_{\text{togg}}(t)=\hat{U}_{\text{rf}}^\dagger(t)\hat{H}_{\text{dd}}\hat{U}_{\text{rf}}(t).
\end{equation}
Here $\hat{U}_{\text{rf}}(t)={\rm e}^{-{\rm i}\mathbf{T}\int_0^t\hat{H}_{\rm rf}^{\rm rot}\dd t}$ is the unitary propagator for the sequence of RF pulses applied up to time $t$. During the $g$-th interval, the rotations induced by the pulses toggle $\hat{H}_{\mathrm{dd}}$ into
\begin{equation}
	\hat{H}_{\text{togg}}^{\{g\}} =\sum_{i<j,\mu,\nu} J_{ij}[3(\hat{\bm{S}}_{i\mu} \cdot \bm{u}_g) (\hat{\bm{S}}_{j\nu} \cdot \bm{u}_g) - \hat{\bm{S}}_{i\mu}\cdot \hat{\bm{S}}_{j\nu}],
    \label{eq:slice}
\end{equation}
where $\bm{u}_g=(u_g^x,u_g^y,u_g^z)$ is a unit vector representing the orientation. Provided that $\hat{U}_{\text{rf}}(t)=\hat{\mathbbm{1}}$ at integer multiples of the cycle time $T$, the dynamics in the toggling frame coincide with those in the original frame under stroboscopic observation. Consequently, the first three terms of the Floquet-Magnus expansion are given in the toggling frame by
\begin{equation}
    \begin{aligned}
        \hat{H}_{\text{F}}^{(0)} &\equiv \frac{1}{T}\sum_{g=1}^{g_{\max}}\hat{H}_{\text{togg}}^{\{g\}}\tau_{g},\\
        \hat{H}_{\text{F}}^{(1)} &\equiv \frac{-\mathrm{i}}{2 T}\sum_{g_1=1}^{g_{\max}}\sum_{g_2=1}^{g_1}\left[\hat{H}_{\text{togg}}^{\{g_1\}}, \hat{H}_{\text{togg}}^{\{g_2\}}\right]\tau_{g_1}\tau_{g_2},\\
        \hat{H}_{\text{F}}^{(2)} &\equiv -\frac{1}{6 T} \sum_{g_1=1}^{g_{\max}}\sum_{g_2=1}^{g_1}\sum_{g_3=1}^{g_2}\left\{\left[\hat{H}_{\text{togg}}^{\{g_1\}},[\hat{H}_{\text{togg}}^{\{g_2\}}, \hat{H}_{\text{togg}}^{\{g_3\}}]\right]+\left[\hat{H}_{\text{togg}}^{\{g_3\}},[\hat{H}_{\text{togg}}^{\{g_2\}}, \hat{H}_{\text{togg}}^{\{g_1\}}]\right]\right\}\tau_{g_1}\tau_{g_2}\tau_{g_3},
    \end{aligned}
    \label{eq:FM_expansion}
\end{equation}
where $T=\sum_{g}\tau_g$ is the total period of the pulse sequence, and $g_{\max}$ is the number of intervals. The norm of the $n$-th order term scales as $\mathrm{O}[(JT)^n]$. If $\hat{H}_{\text{togg}}^{\{g\}}$ during one cycle is symmetric in time (i.e., a palindrome), the odd-order terms of the Floquet-Magnus expansion (Eq.~\eqref{eq:FM_expansion}) vanish, since they involve an odd number of nested commutators. In this case, the total effective Hamiltonian is $\hat{H}=\hat{H}_{\text{F}}^{(0)}+\hat{H}_{\text{F}}^{(2)}+\hat{H}_{\text{F}}^{(4)}+\cdots$.

For the sequence Eq.~\eqref{eq:8pulse}, the zeroth-order average Hamiltonian $\hat{H}_{\text{0th}}$ is the double-quantum-transition Hamiltonian \cite{baum1985multiple}:
\begin{equation}
\hat{H}_{\mathrm{dq}} \equiv \sum\limits_{i<j,\mu,\nu}\frac{J_{ij}}{2}(\hat{S}_{i\mu}^+\hat{S}_{j\nu}^++\hat{S}_{i\mu}^-\hat{S}_{j\nu}^-).
\end{equation}
To suppress all the odd-order terms in the Floquet-Magnus expansion, we symmetrize the 8-pulse sequence to get a 16-pulse sequence
\begin{equation}
	\left(\frac{\tau}{2}, \mathbf{Y}, \tau', \mathbf{Y}, \tau, \mathbf{Y}, \tau', \mathbf{Y}, \tau, \overline{\mathbf{Y}}, \tau', \overline{\mathbf{Y}}, \tau, \overline{\mathbf{Y}}, \tau', \overline{\mathbf{Y}}, \tau,\overline{\mathbf{Y}}, \tau', \overline{\mathbf{Y}}, \tau, \overline{\mathbf{Y}}, \tau', \overline{\mathbf{Y}}, \tau,\mathbf{Y}, \tau', \mathbf{Y}, \tau, \mathbf{Y}, \tau', \mathbf{Y},\frac{\tau}{2}\right).
    \label{eq:seq1}
\end{equation}
The period of this sequence is $T=24(\tau+\tau_{\rm p})=96\ \mu{\rm s}$, so the evolution time for $n$ cycles of the sequence is $t=nT$. Analogously, the reversed Hamiltonian $-\hat{H}_{\rm dq}$ can be realized via the pulse sequence
\begin{equation}
	\left(\frac{\tau}{2}, \mathbf{X}, \tau', \mathbf{X}, \tau, \mathbf{X}, \tau', \mathbf{X}, \tau, \overline{\mathbf{X}}, \tau', \overline{\mathbf{X}}, \tau, \overline{\mathbf{X}}, \tau', \overline{\mathbf{X}}, \tau,\overline{\mathbf{X}}, \tau', \overline{\mathbf{X}}, \tau, \overline{\mathbf{X}}, \tau', \overline{\mathbf{X}}, \tau,\mathbf{X}, \tau', \mathbf{X}, \tau, \mathbf{X}, \tau', \mathbf{X},\frac{\tau}{2}\right).
    \label{eq:seq2}
\end{equation}
The Hamiltonian governing the forward evolution is $\hat{H}=\hat{H}_{\rm dq}+\hat{H}_1$, where $\hat{H}_1=\hat{H}_{\rm 2nd}+\hat{H}_{\rm 4th}+\cdots$; for the backward evolution the total Hamiltonian is $\hat{H}'=-\hat{H}_{\rm dq}-\hat{H}_1'$. The two higher-order contributions differ, $\hat{H}_1'\neq \hat{H}_1$, because the higher-order terms generally cannot be simultaneously reversed.

\subsection{State initialization}
In the absence of RF irradiation, the system resides in a thermal equilibrium state determined by the static magnetic field at room temperature:
\begin{equation}
    \hat{\rho}_0 = \frac{\mathrm{e}^{-\beta\hbar\hat{H}_{\rm lab}}}{\mathcal{Z}} 
    \approx \frac{\mathrm{e}^{-\beta\hbar\omega_{\rm H}\hat{O}_z}}{\mathcal{Z}} 
    \approx \frac{\mathbbm{1} + \epsilon \hat{O}_z}{2^N},
    \label{eq:rho0}
\end{equation}
where $\beta = 1/(k_{\rm B}T_c)$ is the inverse temperature, $\mathcal{Z}$ the partition function, and $\epsilon \equiv -\beta\hbar\omega_{\rm H}$.
The first approximation discards the internal dipolar energies, which are far smaller than the Zeeman term $\hbar\omega_{\rm H}\hat{O}_z$.
The second expands the exponential to linear order, valid under the high-temperature condition $|\epsilon| \ll 1$; at $9.4$~T and room temperature, $|\epsilon| \sim 10^{-5}$.

\subsection{Encoding of phase shift}
The collective rotation $e^{-i\hat{O}_z\phi}$ is implemented by shifting the phases of the RF pulses in the backward evolution sequence:
\begin{equation}
	\left(\frac{\tau}{2}, \mathbf{X}_\phi, \tau', \mathbf{X}_\phi, \tau, \mathbf{X}_\phi, \tau', \mathbf{X}_\phi, \tau, \overline{\mathbf{X}}_\phi, \tau', \overline{\mathbf{X}}_\phi, \tau, \overline{\mathbf{X}}_\phi, \tau', \overline{\mathbf{X}}_\phi, \tau,\overline{\mathbf{X}}_\phi, \tau', \overline{\mathbf{X}}_\phi, \tau, \overline{\mathbf{X}}_\phi, \tau', \overline{\mathbf{X}}_\phi, \tau,\mathbf{X}_\phi, \tau', \mathbf{X}_\phi, \tau, \mathbf{X}_\phi, \tau', \mathbf{X}_\phi,\frac{\tau}{2}\right),
    \label{eq:seq_shift}
\end{equation}
where $\mathbf{X}_\phi$ denotes
\begin{equation}
{\rm e}^{-{\rm i}(\hat{O}_x \cos\phi - \hat{O}_y \sin\phi)\frac{\pi}{2}} = {\rm e}^{{\rm i}\hat{O}_z\phi} {\rm e}^{-{\rm i}\hat{O}_x\frac{\pi}{2}} {\rm e}^{-{\rm i}\hat{O}_z\phi},
\end{equation}
i.e., shifting the rotation axis of the $\pi/2$ pulse from $\hat{x}$ to ($\hat{x}\cos\phi+\hat{y}\sin\phi$). Since $[\hat{O}_z, \hat{H}_{\rm dd}] = 0$, Eq.~\eqref{eq:seq_shift} factorizes as
\begin{equation}
\begin{aligned}
    	&{\rm e}^{{\rm i}\hat{O}_z\phi}\left(\frac{\tau}{2}, \mathbf{X}, \tau', \mathbf{X}, \tau, \mathbf{X}, \tau', \mathbf{X}, \tau, \overline{\mathbf{X}}, \tau', \overline{\mathbf{X}}, \tau, \overline{\mathbf{X}}, \tau', \overline{\mathbf{X}}, \tau,\overline{\mathbf{X}}, \tau', \overline{\mathbf{X}}, \tau, \overline{\mathbf{X}}, \tau', \overline{\mathbf{X}}, \tau,\mathbf{X}, \tau', \mathbf{X}, \tau, \mathbf{X}, \tau', \mathbf{X},\frac{\tau}{2}\right){\rm e}^{-{\rm i}\hat{O}_z\phi}.
\end{aligned}
\end{equation}

\subsection{Signal and readout noise}
\begin{figure}[htb]
    \centering
    \includegraphics[width=0.9\linewidth]{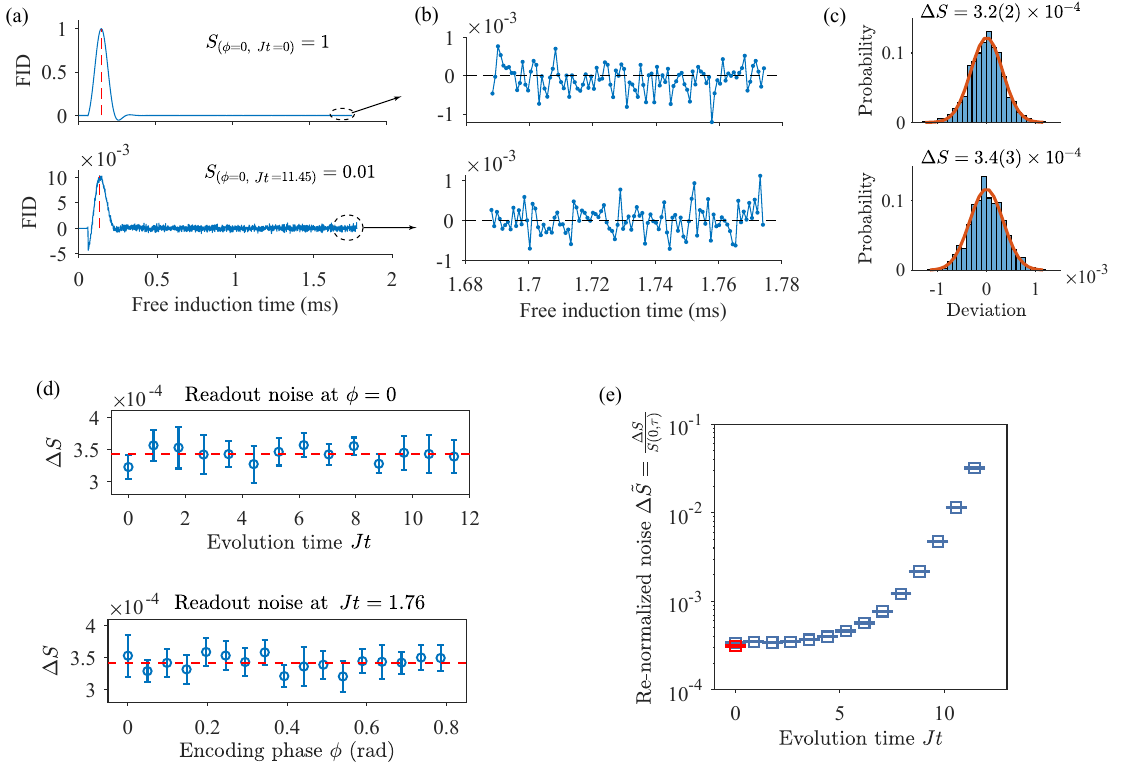}
    \caption{\textbf{FID signal and noise analysis.} (a) FID signals acquired using the magic echo method. The value of $S(\phi,t)$ is determined from the normalized amplitude at the echo peak (indicated by the red dashed line). (b) Exaggeration of the FID tails (last 100 points) in (a). (c) Probability distribution of the random deviations from zero in (b). The statistics are obtained from 10 repeated measurements. Gaussian fits to the distributions yield estimates of the standard deviation $\Delta S$. (d) Dependence of $\Delta S$ on evolution time $Jt$ and encoding phase $\phi$. (e) Growth of the re-normalized readout noise $\Delta \tilde{S}=\Delta S/S(0,t)$ (effectively the noise-to-signal ratio) with $Jt$. For a fixed $Jt$, the standard deviation $\Delta S$ is obtained by pooling all readout noise values across different $\phi$. The initial red point represents the value for the uncorrelated state, i.e., $\Delta\tilde{S}_{\rm (uncorr)}=3.1(1)\times 10^{-4}$.}
    \label{fig:FID}
\end{figure}

The final observable $S(\phi,t)\propto\langle\hat{O}_z\rangle$ and the readout noise $\Delta S$ were determined from the free induction decay (FID). Before acquiring the FID, a waiting period is necessary to allow the excitation circuit ringing to decay. To compensate for signal loss due to evolution under $\hat{H}_{\rm dd}$ during this waiting time, a magic echo pulse sequence was employed. This technique generates a time-reversed dipolar Hamiltonian, $-\hat{H}_{\rm dd}$, which refocuses the magnetization to form an echo~\cite{rhim1970violation}. Additionally, a triple-pulse excitation scheme with phase cycling was implemented to suppress radio-frequency (RF) acoustic ringing artifacts~\cite{wang2021triple}. Two representative FIDs acquired using these methods, for different evolution times $Jt$, are shown in \ref{fig:FID}(a). 

The value of $\langle\hat{O}_z\rangle$ was taken as the amplitude at the peak of the magic echo, indicated by the dashed red line in \ref{fig:FID}(a). The value of $\langle\hat{O}_z\rangle$ at $t=0$ and $\phi=0$ is used as the reference to normalize the signal: $S(\phi,t)=\langle\hat{O}_z\rangle(\phi,t)/\langle\hat{O}_z\rangle(0,0)$. The readout noise is shown in \ref{fig:FID}(b), which exhibits random fluctuations about zero. The probability distribution of the deviations is depicted in \ref{fig:FID}(c), which follows a Gaussian distribution. The standard deviation $\Delta S$ is extracted from Gaussian fits. We further verify that while the signal $S$ depends on $t$ and $\phi$, $\Delta S$ is independent of both $t$ and $\phi$, as demonstrated in \ref{fig:FID}(d). Finally, we plot the re-normalized readout noise $\Delta \tilde{S}=\Delta S/S(0,t)$ in \ref{fig:FID}(e), which essentially corresponds to the noise-to-signal ratio.

\clearpage
\section{Relation between QFI and MQC}
The evolved correlated state $\hat{\rho}_t={\rm e}^{-{\rm i}\hat{H}t}\hat{\rho}_0{\rm e}^{{\rm i}\hat{H}t}$, with $\hat{\rho}_0=\frac{\mathrm{e}^{\epsilon\hat{O}_z}}{\mathcal{Z}} 
    \approx \frac{\mathbbm{1} + \epsilon \hat{O}_z}{2^N}$, can be expanded by a set of Pauli string operators:
\begin{equation}
    \hat{\rho}_t=\frac{\hat{\mathbbm{1}}}{2^N}+\sum_{K=1}^N \sum_{m=-L}^L \sum_{s=1}^{\zeta_{L m}} b_{L m}^s(t) \hat{\mathcal{B}}_{L m}^s.
    \label{eq:rho_tau}
\end{equation}
Here $\hat{\mathcal{B}}_{L m}^s=c_{Lms}\hat{S}_i^\alpha\hat{S}_j^\beta\cdots\hat{S}_k^\gamma$ is a Pauli string operator, with $c_{Lms}$ a normalized coefficient satisfying $\hat{\mathcal{B}}_{L m}^s(\hat{\mathcal{B}}_{L m}^s)^\dagger=\hat{\mathbbm{1}}$, $\alpha,\beta,\gamma\in\{+,-,z\}$ and $i,j,k$ spin indices. $L$ denotes the string length, equal to the number of single-spin operators it contains; $m$ equals the difference between the number of raising and lowering operators, termed the coherence order in the eigenbasis of $\hat{O}_z$; $s$ denotes additional indices needed for complete labeling, and $\zeta_{L m}$ is the number of Pauli strings sharing the same $L$ and $m$. The $m$-th order MQC intensity is $I_m(t)=\frac{1}{\mathcal{C}}\sum_{L \geqslant|m|}^N \sum_{s=1}^{\zeta_{L m}}2^N\left|b_{L m}^s(t)\right|^2$, where $\mathcal{C}$ is a normalization constant ensuring $\sum_m I_m(t)=1$.
An equivalent expression for the MQC intensity, written in the eigenbasis of $\hat{O}_z$, is~\cite{baum1985multiple}
\begin{equation}
    I_m(t)=\frac{1}{\mathcal{C}}\sum_{\substack{k,l \text { s.t. } \\ M_k-M_l=m}}\left|\bra{\psi_k}(\hat{\rho}_t-\hat{\mathbbm{1}} / 2^N)\ket{\psi_l}\right|^2.
    \label{eq:MQC}
\end{equation}
Here, $\{M_k\}$ and $\{\ket{\psi_k}\}$ are the eigenvalues and eigenvectors of $\hat{O}_z$, respectively. The initial state $\hat{\rho}_0=e^{\epsilon\hat{O}_z}/\mathcal{Z}$ contains no quantum coherences ($m=0$). During the forward evolution, MQCs are generated following the selection rule $\Delta m=\pm 2$. The spin operators respond to a global rotation ${\rm e}^{-{\rm i}\hat{O}_z\phi}$ according to their coherence order:
\begin{equation}
{\rm e}^{-{\rm i}\hat{O}_z\phi}\hat{\mathcal{B}}_{L m}^s{\rm e}^{{\rm i}\hat{O}_z\phi}=\mathrm{e}^{-\mathrm{i} m \phi}\hat{\mathcal{B}}_{L m}^s,
\end{equation}
which reflects the fact that an $m$-th order quantum coherence amplifies the phase by a factor of $m$ compared to single-quantum coherences.

For a general mixed probe state with eigendecomposition $\hat{\rho}=\sum_{k}p_k\ket{\psi_k}\bra{\psi_k}$, the QFI is given by~\cite{braunstein1994statistical,helstrom1969quantum}
\begin{equation}
\mathcal{F}_{\rm Q}(\hat{\rho},\hat{O})\equiv 2\sum_{k,l}\frac{\left(p_k-p_l\right)^2}{p_k+p_l}|\bra{\psi_k}\hat{O}\ket{\psi_l}|^2,
\label{eq:QFI_raw}
\end{equation}
where $\hat{O}\equiv{\rm i}\hat{U}_\alpha^\dagger(\partial\hat{U}_\alpha/\partial \alpha)$ is the generator of the $\alpha$-dependent unitary encoding $\hat{U}_\alpha$. In our case, the phase-shift encoding is $\hat{U}_\phi={\rm e}^{-{\rm i}\hat{O}_z\phi}$ and the generator is the collective spin operator $\hat{O}_z$. For the probe state $\hat{\rho}_t$ given by Eq.~\eqref{eq:rho_tau}, unitary evolution preserves the eigendecomposition, so the eigenvalues of $\hat{\rho}_t$ are inherited from the initial state $\hat{\rho}_0=e^{\epsilon\hat{O}_z}/\mathcal{Z}$. We thus have 
\begin{equation}
\begin{aligned}
    p_k(t)=&p_k(0)=e^{\epsilon M_k}/\mathcal{Z},\\
    \ket{\psi_k(t)}&={\rm e}^{-{\rm i}\hat{H}t}\ket{\psi_k(0)}.
\end{aligned}
\end{equation}
Substituting these into Eq.~\eqref{eq:QFI_raw}, we obtain 
\begin{equation}
\begin{aligned}
    \mathcal{F}_{\rm Q}(\hat{\rho}_t,\hat{O}_z)
    &=2\sum_{k,l}\left(\frac{e^{\epsilon M_k}-e^{\epsilon M_l}}{e^{\epsilon M_k}+e^{\epsilon M_l}}\right)^2(p_k+p_l)\abs{\bra{\psi_k}e^{i\hat{H}t}\hat{O}_ze^{-i\hat{H}t}\ket{\psi_l}}^2\\
    &=2\sum_{k,l}\tanh^2\left[\frac{\epsilon(M_k-M_l)}{2}\right](p_k+p_l)\abs{\bra{\psi_k}\hat{O}_z(-t)\ket{\psi_l}}^2,
\end{aligned}
\end{equation}
which reproduces the result derived in Ref.~\onlinecite{hauke2016measuring} for evaluating the QFI of probe states with a thermal spectrum. 

In our case, owing to the finite size of the correlated clusters (which corresponds to the support size of $\hat{O}_z(-t)$), the total $N$ spins can be effectively decomposed into $N/K$ disentangled $K$-spin clusters, yielding $\mathcal{F}_{\rm Q}=(N/K)\mathcal{F}_{\rm Q}^{(K)}$. To calculate the QFI for a single $K$-spin cluster, $\mathcal{F}_{\rm Q}^{(K)}$, we apply the condition $\epsilon K\ll 1$. This condition is well satisfied since $\epsilon\sim 10^{-5}$ and $K(t)\lesssim 10^4$ (with the fitting results for $K(t)$ detailed in \ref{sec:fit k}), allowing us to approximate
\begin{equation}
    p_k+p_l=\frac{e^{\epsilon M_k}+e^{\epsilon M_l}}{\mathcal{Z}}\approx \frac{2+\epsilon( M_k+M_l)}{2^K}=\frac{2+\mathrm{O}(K\epsilon)}{2^K},
\end{equation}
leading to
\begin{equation}
\begin{aligned}
\mathcal{F}_{\rm Q}^{(K)}=&\frac{4+\mathrm{O}(K\epsilon)}{2^K}\sum_{m=-K}^{K}\left(\tanh{\frac{\epsilon m}{2}}\right)^2\sum_{\substack{k,l \text { s.t. } \\ M_k-M_l=m}}|\bra{\psi_k}\hat{O}_z(-t)\ket{\psi_l}|^2\\
=&\frac{4+\mathrm{O}(K\epsilon)}{2^K}[\Tr_K(\hat{O}_z^2)]\sum_{m=-K}^{K} \left(\tanh{\frac{\epsilon m}{2}}\right)^2 I_m(t)\\
=&[1+\mathrm{O}(K\epsilon)] K\sum_{m=-K}^{K}\left(\tanh{\frac{\epsilon m}{2}}\right)^2I_m(t).
\end{aligned}
\end{equation}
These intermediate steps follow from the definition of the MQC intensity in Eq.~\eqref{eq:MQC}, the trace identity $\Tr_K(\hat{O}_z^2)=K2^K/4$, and the symmetry $I_m(t)=I_m(-t)$. Furthermore, since $\epsilon K\ll 1$, we can approximate $\tanh\left(\frac{\epsilon m}{2}\right)\approx\frac{\epsilon m}{2}$, which ultimately yields
\begin{equation}
\begin{aligned}
    \mathcal{F}_{\rm Q}(\hat{\rho}_t,\hat{O}_z)=& \frac{N}{K}\mathcal{F}_{\rm Q}^{(K)}\\
    =&[1+\mathrm{O}(K\epsilon)]\frac{N}{4}\epsilon^2 \sum_m m^2 I_m(t).
    \end{aligned}
\end{equation}

\section{Gaussian fit of the MQC spectrum} 
\label{sec:fit k}
To compute the QFI, the summation over $m$ often cannot be evaluated directly because higher-order MQCs are difficult to measure precisely due to the finite signal-to-noise ratio. Following standard practice, the distribution of $I_m$ is modeled by a Gaussian profile $I_m\propto e^{-m^2/K}$, where the variance $K=2\sum_m m^2 I_m$ quantifies the effective number of correlated spins~\cite{alvarez2015localization}. Under this approximation, the second moment ($K/2$) can be determined by fitting the MQC spectrum.

\begin{figure}
    \centering
    \includegraphics[width=0.8\linewidth]{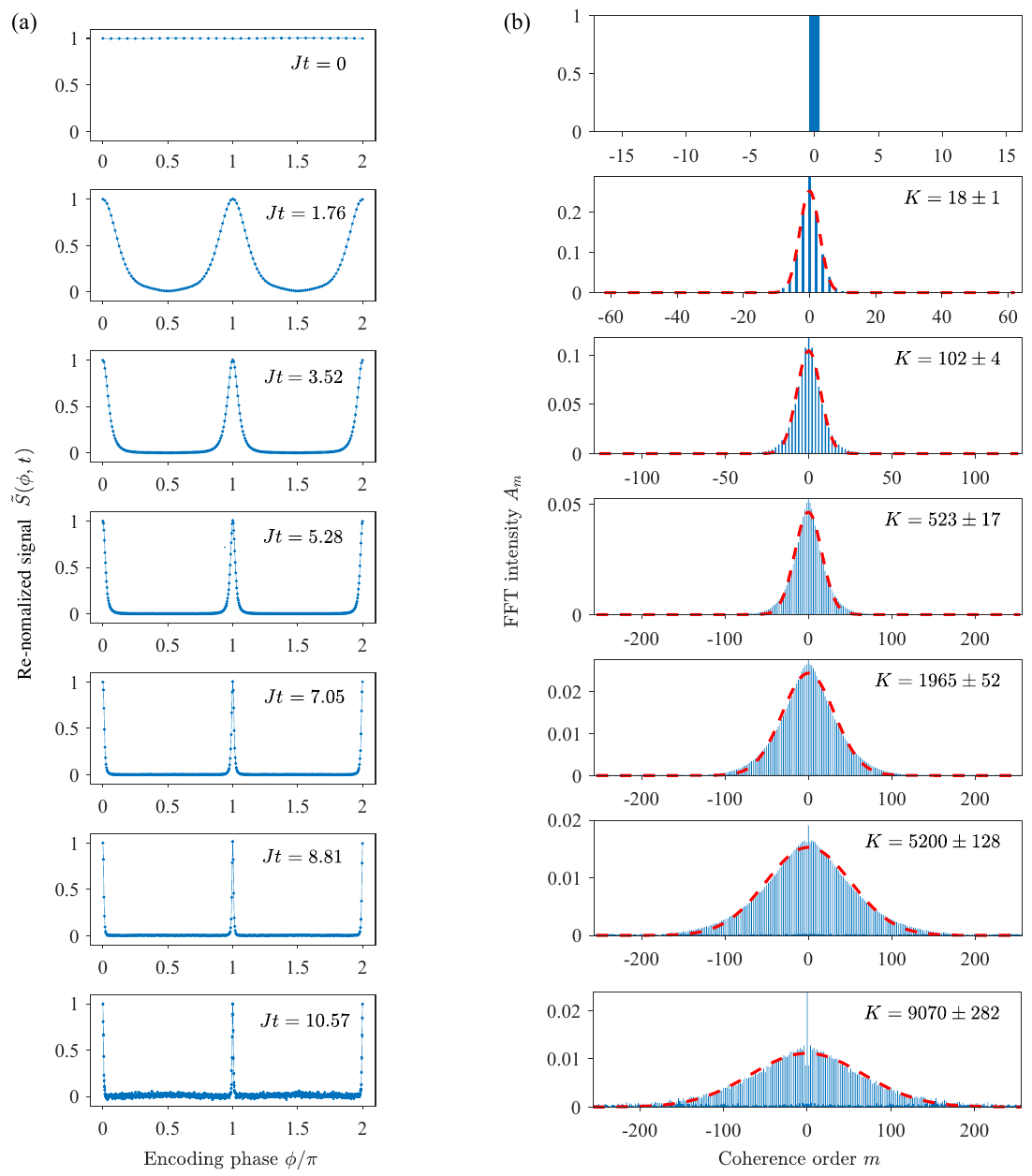}
    \caption{Re-normalized signal $\tilde{S}(\phi,t) = S(\phi, t)/S(0,t)$ and Gaussian fits to the FFT spectrum $\{A_m\}$. In panel (a), the re-normalized readout noise $\Delta \tilde{S}$ ranges from $10^{-4}$ to $10^{-2}$ as $Jt$ increases.}
    \label{fig:Gaussian fit}
\end{figure}

As mentioned in the main text, due to imperfect time reversal, the normalization condition $\sum_m I_m(t)=1$ cannot be maintained with increasing evolution time $t$ in the experiment. A straightforward remedy is to re-normalize the signal at each evolution time. The re-normalized signal is $\tilde{S}(\phi,t) = S(\phi, t)/S(0,t)=\sum_m A_m(t) \mathrm{e}^{-\mathrm{i} m \phi}$, which ensures $\sum_m A_m(t) = 1$. The coefficients $A_m(t)$ correspond to the MQC intensities $I_m(t)$ in the ideal limit $\delta\hat{H} = 0$. By sampling $\phi$ from $0$ to $2\pi$ and performing a discrete Fourier transform of $\tilde{S}(\phi,t)$ with respect to $\phi$, the spectrum $A_m(t)$ is extracted. The results of $\tilde{S}(\phi,t)$ at various evolution times $t$, along with the corresponding $A_m(t)$ spectra, are shown in \ref{fig:Gaussian fit}. Gaussian fits of the form $A_m(t)\propto e^{-m^2/K}$ and the fitted values of $K(t)$ are also presented.

\clearpage
\section{Phase sensitivity for the uncorrelated state}
\begin{figure}
    \centering
    \includegraphics[width=0.3\linewidth]{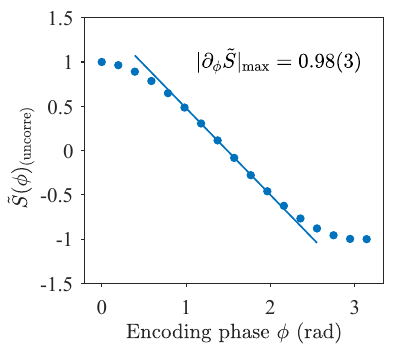}
    \caption{Signal response of the uncorrelated state $\hat{\rho}_{\rm (uncorr)}\propto \mathbbm{1}+\epsilon \hat{O}_x$ to a global rotation $e^{-i\hat{O}_z\phi}$. The observable is $\langle\hat{O}_x\rangle$. Circles: experimental data; Line: linear fit. The error bars are contained within the data markers.}
    \label{fig:uncorre}
\end{figure}

As a reference, we prepare the uncorrelated state $\hat{\rho}_{\rm (uncorr)}\propto \mathbbm{1}+\epsilon \hat{O}_x$, apply the global rotation ${\rm e}^{-{\rm i}\hat{O}_z\phi}$, and measure $\langle\hat{O}_x\rangle$. Ideally, this yields a signal $\tilde{S}_{\rm (uncorr)}=\cos \phi$, as shown in \ref{fig:uncorre}. Via linear fits using a 5-point sliding window, we calibrate the maximum slope $\tilde{g}_{(\rm uncorre)}=\max_\phi|\partial_\phi\tilde{S}_{\rm (uncorr)}|=0.98(3)$. Therefore, given a readout noise of $\Delta\tilde{S}_{\rm (uncorr)}=3.1(1)\times 10^{-4}$ [red point in \ref{fig:FID}(e)], the phase sensitivity is $\Delta\phi_{\rm (uncorr)}=\Delta\tilde{S}_{\rm (uncorr)}/\tilde{g}_{(\rm uncorr)}=321(11)\ \mu$rad.

\bibliography{ref}